\documentstyle[aps,preprint,psfig]{revtex}

\tighten

\newcommand{\epem}{e^+e^-}
\newcommand{\emem}{e^-e^-}
\newcommand{\selectron}{\tilde{e}}
\newcommand{\sneutrino}{\tilde{\nu}}
\newcommand{\neutralino}{\tilde{\chi}^0}
\newcommand{\LSP}{\tilde{\chi}^0_1}
\newcommand{\neutralinotwo}{\tilde{\chi}^0_2}
\newcommand{\chargino}{\tilde{\chi}^{\pm}}
\newcommand{\charginoone}{\tilde{\chi}^{\pm}_1}
\newcommand{\gluino}{\tilde{g}}

\newcommand{\susyU}{\widetilde{U}}
\newcommand{\dsusyU}{\widetilde{U}}

\begin{document}

\draft
\pagestyle{empty}

\preprint{
\noindent
\hfill
\begin{minipage}[t]{3in}
\begin{flushright}
FERMILAB--PUB--97/205--T \\
LBNL--40466 \\
UCB--PTH--97/34 \\
RU--97--46 \\
hep-ph/9706476 \\
June 1997
\end{flushright}
\end{minipage}
}

\title{Signatures of multi-TeV scale particles\\
 in supersymmetric theories
}

\author{Hsin-Chia Cheng$^1$, Jonathan L. Feng$^2$
\thanks{Research Fellow, Miller Institute for Basic Research in Science.}, 
and Nir Polonsky$^3$}

\address{\vspace*{0.2in}
$^1$ Fermi National Accelerator Laboratory \\
P.~O.~Box 500, Batavia, Illinois 60510}

\address{\vspace*{0.07in}
$^2$ Theoretical Physics Group, Lawrence Berkeley
National Laboratory \\ 
and Department of Physics, University of
California, Berkeley, CA 94720}

\address{\vspace*{0.07in}
$^3$ Department of Physics and Astronomy \\ 
Rutgers University, Piscataway, NJ 08855-0849 } 

\maketitle

\begin{abstract}
Supersymmetric particles at the multi-TeV scale will escape direct
detection at planned future colliders. However, such particles induce
non-decoupling corrections in processes involving the accessible
superparticles through violations of the supersymmetric equivalence
between gauge boson and gaugino couplings.  In a previous study, we
parametrized these violations in terms of super-oblique parameters and
found significant deviations in well-motivated models. Here, we
systematically classify the possible experimental probes of such
deviations, and present detailed investigations of representative
observables available at a future linear collider.  In some scenarios,
the $e^-e^-$ option and adjustable beam energy are exploited to
achieve high precision.  It is shown that precision measurements are
possible for each of the three coupling relations, leading to
significant bounds on the masses and properties of heavy
superparticles and possible exotic sectors.
\end{abstract}


\pagestyle{plain}

\section{Introduction}
\label{sec:introduction}

If supersymmetry (SUSY) has relevance for the gauge hierarchy problem,
fine-tuning considerations~\cite{finetuning} suggest that
supersymmetric particles typically have mass on order of or below the
TeV scale.  The discovery of some supersymmetric particles is
therefore eagerly anticipated at present and future colliders.  In
particular, the Large Hadron Collider (LHC)~\cite{LHC} at CERN is
likely to discover squarks and gluinos up to masses of $1-2 \text{
TeV}$~\cite{SnowmassSUSY,SnowmassLHC,LHCsquarks}, and proposed linear
$\epem$ colliders~\cite{JLC,NLC,DESYLC}, with $\sqrt{s} = 0.5 - 1.5
\text{ TeV}$, will be able to discover pair-produced superpartners with
masses close to the kinematic
limit~\cite{SnowmassSUSY,JLC,MP,SnowmassNLC}.

It is possible, however, that some number of the superpartners of the
standard model (SM) particles are heavy and beyond the discovery reach
of planned future colliders. In fact, as will be described in more
detail below, a wide variety of models predict superparticle spectra
leading to such scenarios. If this possibility is actually realized in
nature, we must then rely solely on indirect methods to probe the
masses and properties of these heavy superparticles, at least until
colliders at even higher energies become available.  In most
experimentally accessible processes, heavy supersymmetric states
decouple, and their effects are not measurable for the large masses
we are considering. However, the larger these masses are, the more
they break SUSY, and so their effects may appear at detectable levels
in processes involving light superpartners as violations of hard
supersymmetric relations, {\em i.e.}, supersymmetric relations between
dimensionless coupling constants.  For example, consider the gauge
couplings $g_i$, where the subscript $i=1, 2, 3$ refers to the U(1),
SU(2), or SU(3) gauge group, and their SUSY counterparts, the
gaugino-fermion-sfermion couplings, which we denote by $h_i$.  In the
limit of unbroken supersymmetry,

\begin{equation}
g_i =h_i \ .
\label{equivalence}
\end{equation}
However, the large SUSY breaking masses of heavy superpartners lead to
deviations from these SUSY relations in the low energy effective
theory where the heavy superpartners are decoupled.  These deviations
are non-decoupling and grow logarithmically with the heavy
superpartner masses.  In addition, Eq.~(\ref{equivalence}) is
model-independent and valid to all orders in the limit of unbroken
SUSY.  Deviations from the relations of Eq.~(\ref{equivalence}) are
therefore unambiguous signals of SUSY breaking mass splittings.  Thus,
the masses of kinematically inaccessible sparticles may be measured by
precise determinations of such deviations from processes involving the
accessible sparticles.

The corrections to Eq.~(\ref{equivalence}) from split supermultiplets
are very similar to the oblique corrections~\cite{Peskin,oblique} from
split SU(2) multiplets in the standard model.  This analogy has been
described in detail in a previous paper~\cite{CFP} and was noted in
Ref.~\cite{LR}.  Ignoring Yukawa couplings, these corrections are
dominantly from differences in the wavefunction renormalizations of
gauge bosons and gauginos, which result from inequivalent loops after
the decoupling of heavy superpartners.  Such corrections are therefore
most similar to those described by the $U$ parameter of the oblique
corrections~\cite{Peskin}, which is a measure of the difference
between the wavefunction renormalizations of the $W$ and $Z$ gauge
bosons arising from custodial isospin breaking masses in SU(2)
multiplets. For this reason, in Ref.~\cite{CFP} we called the
corresponding SUSY corrections ``super-oblique corrections'' and
defined a set of ``super-oblique parameters,'' $\susyU_i$, one for
each gauge group, which measure deviations from
Eq.~(\ref{equivalence}).  These parameters are given by~\cite{CFP}

\begin{equation}
\susyU_{i} \equiv \frac{h_{i}(m)}{g_{i}(m)} - 1 
\approx \frac{g_{i}^{2}(m)}{16\pi^{2}}(b_{g_{i}} - b_{h_{i}})
\ln \frac{M}{m} \ ,
\label{delta}
\end{equation}
where $M (m)$ is the heavy (light) superpartner scale, and $b_{g_i}
(b_{h_i})$ is the one-loop $\beta$-function coefficient for the gauge
(gaugino) coupling in the effective theory between the heavy and light
mass scales. Note that $b_{g_i} > b_{h_i}$, and so the super-oblique
parameters are always positive (at the leading logarithm
level)~\cite{CFP}.  We also defined two-index parameters measuring the
relative deviations of two gauge groups,

\begin{equation}
\label{rho}
\dsusyU_{ij} \equiv \frac{h_{i}(m)/h_{j}(m)}{g_{i}(m)/g_{j}(m)} - 1
 \approx \susyU_i - \susyU_j \ .
\end{equation}
The parameters $\dsusyU_{ij}$ are simple linear combinations of the
$\dsusyU_i$, but are physically relevant, as they are quantities that
may be probed in branching ratio measurements, as we will see in an
example below.  These super-oblique parameters parametrize universal
effects that enter all processes involving gaugino-fermion-sfermion
interactions, and their simple form allows us to study such
non-decoupling effects in a model-independent fashion. Other
flavor-dependent non-decoupling corrections, for example, those
induced by Yukawa couplings, and additional super-oblique corrections
$\widetilde{T}_i$ were also described in Ref.~\cite{CFP}; we refer
interested readers to that study for discussion of these and other
issues.

Depending on which superpartners are heavy, the models that contain
heavy superparticles may be roughly divided into two
categories~\cite{CFP}: ``heavy QCD models'' and ``2--1 models.'' In
heavy QCD models, all strongly-interacting superpartners, {\em i.e.},
the gluino and all squarks, are in the heavy sector. Their large SUSY
breaking masses may arise from either the proportionality of soft
masses to standard model gauge coupling constants or the
renormalization group evolution effects of a large gluino mass.
Examples of such models include the no-scale limit of minimal
supergravity~\cite{noscale}, models of gauge-mediated SUSY
breaking~\cite{gm}, and models with non-universal gaugino masses and a
heavy gluino~\cite{ACM}. The super-oblique corrections in these models
have been calculated in Ref.~\cite{CFP}, and the results are

\begin{eqnarray}
\label{delta2a}
\susyU_{2} &\approx& 0.80\% \times \ln R \ , \\
\label{delta1a}
\susyU_{1} &\approx&  0.29\% \times \ln R \ , \\
\label{rho21a}
\dsusyU_{21} &\approx&  0.50\% \times \ln R \ , 
\end{eqnarray}
where $R=M/m$ is typically ${\cal O}(10)$ in heavy QCD models.

In 2--1 models, the scalars of the first two generations are heavy and
the third generation scalars are at the weak
scale~\cite{effectivesusy}.  These models are motivated by attempts to
solve the SUSY flavor problem with heavy first two generation scalars
while avoiding extreme fine-tuning problems by keeping the third
generation scalars, which couple strongly to the Higgs sector, at the
weak scale. Assuming all gauginos to be in the light sector, the
super-oblique corrections in 2--1 models were found in Ref.~\cite{CFP}
to be

\begin{eqnarray}
\label{delta3b}
\susyU_{3} &\approx& 2.5\% \times \ln R \ , \\
\label{delta2b}
\susyU_{2} &\approx&  0.71\% \times \ln R \ , \\
\label{delta1b}
\susyU_{1} &\approx&  0.35\% \times \ln R \ , \\
\label{rho32}
\dsusyU_{32} &\approx& 1.8\% \times \ln R \ , \\
\label{rho31}
\dsusyU_{31} &\approx&  2.2\% \times \ln R \ , \\
\label{rho21b}
\dsusyU_{21} &\approx&  0.35\% \times \ln R \ .
\end{eqnarray}
In 2--1 models, values of $R$ in the range $\sim 40-200$ may be taken
as typical.

Although the values of expected super-oblique parameters vary from
model to model, they are always proportional to the square of their
corresponding standard model gauge couplings, as is clear from
Eq.~(\ref{delta}).  Thus, we typically expect the parameters
$\susyU_3$, $\dsusyU_{31}$, and $\dsusyU_{32}$ to be the largest, and,
for example, a 1\% measurement of $\susyU_2$ is more powerful than a
1\% measurement of $\susyU_1$ for the purposes of bounding new physics
scales.  Finally, note that extra vector-like fields with both SUSY
preserving and SUSY breaking masses, such as the messengers in gauge
mediation models, may also contribute to the super-oblique parameters.
Such contributions were also calculated in Ref.~\cite{CFP}, and were
found to be typically small, with significant contributions only for
very highly split supermultiplets.

The possibility of measuring the supersymmetric couplings $h_i$ and
testing the relations $g_i = h_i$ has been discussed previously. In
the original proposal~\cite{FMPT}, the possibility of testing the
SU(2) relation through chargino production at the Next Linear Collider
(NLC) was explored. Here the focus was on establishing the identity of
new particles as superpartners through the verification of SUSY
relations. A test of the U(1) relation through $\epem \to
\selectron^+_R \selectron^-_R$ was considered in Ref.~\cite{slepton}.
In this study, both the possibilities of verifying SUSY relations and
of being sensitive to deviations arising from heavy sparticle
thresholds were considered.  Corrections to hard supersymmetry
relations were previously studied in Ref.~\cite{Hikasa}, where
deviations in squark widths were calculated.  However, the possibility
of experimentally verifying such deviations was not addressed.

In this paper, we will systematically classify the many experimental
observables that depend on the couplings $h_i$ and are therefore
formally candidates for measuring super-oblique parameters.  We then
consider three representative examples of observables that may be
sufficiently sensitive to such parameters to yield interesting
results.  Even after including many experimental errors and the
theoretical uncertainties arising from the plethora of unknown SUSY
parameters, we find some promising prospects for very high precision
measurements.  The results have implications for collider design, as
certain options, particularly the $e^-e^-$ mode and adjustable beam
energies, will be seen to be particularly useful.  It is important to
note that a complete study will require detailed experimental
simulations appropriate to the particular scenario realized in nature,
and the case studies we consider typically require measurements beyond
the first stage of experimental study.  However, given that the
measurements discussed here may be the only experimental window on
physics beyond the TeV scale for the foreseeable future, such issues
are well worth investigation.

We begin in Sec.~\ref{sec:observables} by identifying the many
experimental observables that may possibly be used to detect
variations in the hard SUSY relations.  Of course, not all of these
observables may be measured precisely enough to provide significant
bounds on heavy superpartner masses.  In Sec.~\ref{sec:uncertainties}
we discuss the many uncertainties, both experimental and theoretical,
that appear in any measurement, and we describe our treatment of these
errors.  In Secs.~\ref{sec:charginos}--\ref{sec:squarks}, detailed
discussions of the precisions achievable are given for three
representative examples, one for each coupling constant relation.  In
Sec.~\ref{sec:charginos}, we will find that chargino production at the
NLC gives bounds on the heavy mass scale comparable to those achieved
from $\epem \to \selectron_R^+ \selectron_R^-$ in Ref.~\cite{slepton}.
In Sec.~\ref{sec:selectrons}, we improve upon both of these results by
considering selectron production in the $\emem$ mode, where a number
of beautiful properties may be exploited to reach very high precision.
Finally, in Sec.~\ref{sec:squarks}, we find that significant
constraints on the SU(3) super-oblique parameter may also be possible
from squark branching ratios in particular regions of parameter space.
These examples are by no means exhaustive.  However, they make use of
three different sets of sparticles, and are presented to emphasize the
variety of precise probes that may be used to provide interesting
bounds.  The numerous implications of such measurements are collected
in Sec.~\ref{sec:conclusions}.

\section{Observable Probes of Super-oblique Corrections}
\label{sec:observables}

As seen in the previous section, heavy superpartners may induce
significant corrections to all three coupling constant relations $g_i
= h_i$.  We now discuss what observables at colliders have dependences
on the couplings $h_i$ and are therefore candidates for testing these
relations and determining the super-oblique parameters. In this
section, we will concentrate on measuring the couplings $h_i$ at the
light superparticle mass scale $m$.  Such measurements allow one to
measure the heavy sparticle mass scale $M$.  Of course, measurements
of $h_i$ at higher momentum transfers $p^2 > m^2$ may also be
extremely useful, and would allow one to verify the convergence of
$\susyU_i \to 0$ as $p^2 \to M^2$.\footnote{We thank X.~Tata for this
proposal.} Here, however, we will focus on the classification and
measurement of observables at $p^{2} = m^{2}$, leaving the latter for
future studies.  We begin with observables at $\epem$ (and $\emem$)
colliders, where the ability to make precise model-independent
measurements of a wide variety of SUSY parameters is most promising.
The $\epem$ observables all have analogues at hadron colliders, and we
then turn to hadron colliders and discuss briefly which of these
appear most promising in that experimental environment.  Analogous
observables may also be found at a $\mu^+\mu^-$ collider, with
appropriate and obvious replacements of selectrons by smuons in the
case of electroweak observables.

\subsection{Observables at $\protect\bbox{\epem}$ Colliders}
\label{subsec:epem}

Each kinematically accessible superpartner brings with it a set of
observables.  We consider each superpartner in turn, grouping together
those that are similar for this analysis.
  
\subsubsection{Charginos and Neutralinos}

If charginos are kinematically accessible, their production cross
sections are possible probes.  This applies formally to all reactions,
ranging from chargino pair production to more unusual processes where
charginos are produced in association with other particles, such as in
$\chargino e^{\mp} \sneutrino$ production. In the most obvious and
useful example, charginos are pair-produced in $\epem$ collisions
through $s$-channel $\gamma$ and $Z$ diagrams and $t$-channel
sneutrino exchange.  The latter diagram depends on the coupling $h_2$,
and so chargino pair production cross sections may be used to measure
the parameter $\susyU_2$.  In fact, this will serve as our first
example in Sec.~\ref{sec:charginos}.  If charginos have two or more
open decay modes, their branching fractions may be also be
used.\footnote{Of course, individual decay widths may also depend on
the couplings $h_i$.  In special circumstances, such as when the
decays are extremely suppressed and the decay lengths are macroscopic,
the widths themselves may be measurable.  In general, however,
individual decay widths are very difficult to measure, and we will
therefore concentrate on their ratios in the following.}  For example,
if decays $\chargino \to \tilde{f} f'$ and $\chargino \to W^\pm
\neutralino$ are both open, the ratio of these branching fractions is
dependent on $h_2^2/g_2^2$ (if the chargino is pure Wino) and may
serve as a probe as well.

For neutralinos, the situation is similar.  Neutralino pair production
cross sections depend on $h_1$ and $h_2$ through diagrams with
$t$-channel $\tilde{e}$ exchange.  Their branching fractions are also
accessible probes when two or more decay modes are competitive.

An interesting effect of the super-oblique corrections for charginos
and neutralinos is the modification of their mass matrices. For
example, the conventional chargino mass terms are $(\psi ^-)^T
\bbox{M_{\chargino}} \psi^+ + {\rm h.c.}$, where $(\psi^{\pm})^T =
(-i\tilde{W}^{\pm}, \tilde{H}^{\pm})$ and
                                          
\begin{equation}
\bbox{M_{\chargino}} = \left( \begin{array}{cc}
 M_2                    &\sqrt{2} \, m_W \sin\beta \\
\sqrt{2} \, m_W\cos\beta   &\mu              \end{array} \right) \ .
\end{equation}
Here $M_2$ is the SU(2) gaugino mass, $\tan\beta$ is the ratio of
Higgs vacuum expectation values, and $\mu$ is the Higgsino mass
parameter.  The off-diagonal entries of the mass matrix result from
the interactions $H \tilde{W} \tilde{H}$.  In the presence of
super-oblique corrections, these entries must be modified by $m_W
\to (h_2/g_2) m_W$.  Similar comments apply to the neutralino
mixing matrix.  Thus, precise measurements of the chargino and
neutralino masses and mixings may also yield bounds on the
super-oblique parameters.  Such precision measurements were in fact
studied for charginos in Ref.~\cite{FMPT}.  In the mixed region, where
there is large gaugino-Higgsino mixing, interesting bounds may be
obtained, although measurements of the super-oblique parameters at the
percent level appear difficult.  However, in the regions of parameter
space in which charginos and neutralinos are nearly pure gauginos or
Higgsinos, the dependence on the off-diagonal terms is small, and the
effects of super-oblique parameters through the mass matrices are
negligible.

Before considering other sparticles, a few comments are in order.
First, it is clear that no tests are applicable in all regions of
parameter space.  For the observables above to be sensitive to the
super-oblique parameters, for example, it is necessary not only that
charginos and neutralinos be produced, but also that they have either
large gaugino components or substantial gaugino-Higgsino mixing.
Second, all observables depend on many additional SUSY parameters,
including, for example, the masses and compositions of the charginos
and neutralinos, and the masses of the sfermions entering the process.
Thus, a determination of $h_i$ requires a simultaneous determination
of many other parameters.  This is one of the essential difficulties
in these analyses, and will be addressed in detail in the case studies
of the following sections.

\subsubsection{First Generation Sleptons}

For measurements of super-oblique parameters, selectrons
$\tilde{e}_{L,R}$ and electron sneutrinos $\tilde{\nu}_e$ afford
special opportunities.  For example, selectron pair-production cross
sections receive contributions from $t$-channel neutralino exchange,
and so the $\tilde{e}_R \tilde{e}_R$ and $\tilde{e}_R \tilde{e}_L$
cross sections depend on $h_1$, while the $\tilde{e}_L \tilde{e}_L$
cross section depends on both $h_1$ and $h_2$.  This dependence was
exploited in Ref.~\cite{slepton} to measure $h_1$ at $\epem$
colliders.  Note, however, that selectrons, unlike gauginos, may also
be produced in pairs in $\emem$ collisions.  Such reactions may lead
to particularly precise measurements and will be discussed in detail
in Sec.~\ref{sec:selectrons}.  Selectron branching fractions may also
be useful when two decay modes are open.  For example, the ratio
$B(\selectron_L \to e \tilde{W}) / B(\selectron_L \to e \tilde{B})$
depends on $h_2^2/h_1^2$, and may therefore be used to probe
$\dsusyU_{21}$.

Electron sneutrinos may also be produced in $\epem$ collisions.  Their
production cross sections receive contributions from $t$-channel
chargino exchange, and so are sensitive to $h_2$.  Their branching
ratios may also be used.

\subsubsection{Squarks, Gluinos, Higgses, and Other Sleptons}

If gluinos and the other scalars (squarks, Higgs bosons, and second or
third generation sleptons) are accessible, they may also provide
useful information.  Cross sections for production in association with
gauginos, for example, $\sigma (\epem \to \tilde{q}\bar{q}\gluino)$,
depend on $h_i$ couplings.  In addition, as with the other particles,
their branching ratios are also possible probes.  We will consider the
case of squark branching ratios in Sec.~\ref{sec:squarks}.

\subsection{Observables at Hadron Colliders}

All of the observables mentioned above have analogues at hadron
colliders.  A promising aspect of hadron colliders is that strongly
interacting sparticles may be produced in great numbers, allowing
probes of the QCD relations, where the greatest deviations are
expected. The production cross sections of gluinos and squarks are
dependent on the couplings $h_i$.  Unfortunately, cross section
measurements at hadron colliders are open to systematic uncertainties
that, at the level of precision we require for this study, make such
measurements rather difficult.  On the other hand, branching ratios
may be well measured. For example, if squarks may decay to both
gluinos and electroweak gauginos, the relative rates may be a
sensitive probe of the super-oblique corrections.  Similar comments
apply to sleptons and electroweak gauginos when more than one decay
path is open.  The extent to which these branching ratios may be
measured depends strongly on the efficiency for extracting these
signals from background, and is dependent on many SUSY parameters.  In
this study, we will concentrate on $\epem$ probes, although, given the
exciting prospects for discovering SUSY at the LHC, probes there
certainly merit attention, especially if portions of the sparticle
spectrum are not observed or branching ratios deviate from
expectations.

\subsection{Probes of other non-decoupling corrections}
\label{subsec:other}

So far we have concentrated on observables involving gaugino
interactions as probes of the super-oblique corrections.  In fact,
however, heavy superpartner sectors may also induce non-decoupling
effects in interactions that do not involve gauginos.  In particular,
as discussed in Ref.~\cite{CFP}, $D$-term quartic scalar couplings
also receive corrections.  Such corrections appear in a wide variety
of observables.  

Nevertheless, they are generically highly challenging to probe
experimentally.  To begin with, the couplings of four physical scalars
are extremely difficult to measure.  However, $D$-term couplings also
result in cubic scalar couplings when one field is a Higgs.  These
appear in more accessible observables, including, for example, the
widths of heavy Higgs boson decays to sfermions $H, A \to \tilde{f}
\tilde{f}^*$ and $H^{\pm} \to \tilde{f}\tilde{f}'$.  (Note that the 
$D$-term trilinear terms discussed here involve same-chirality
sfermions and are not suppressed by Yukawa couplings; they may thereby
be distinguished from Yukawa-suppressed trilinear terms that originate
from $F$-terms or from soft SUSY breaking trilinear interactions.)
Unfortunately, in the models we are considering, heavy Higgs bosons
may be very heavy, since their mass is governed by $\mu$, which, given
the constraint of the $Z$ boson mass, is typically at the third
generation squark mass scale.  In addition, heavy Higgs bosons are
difficult to study at hadron colliders, and their interactions depend
on a number of other parameters, such as $\tan\beta$ and the $CP$-even
Higgs mixing angle $\alpha$.  Finally, $D$-terms contribute to SU(2)
doublet mass splittings, such as the splitting between
$m_{\selectron_L}$ and $m_{\sneutrino_e}$.  However, these
contributions are only small fractional deviations in already small
mass splittings.  In summary, the $D$-term non-decoupling effects may
be relevant in certain scenarios, for example, if a heavy Higgs is
accessible at an $\epem$ collider.  However, they do not generally
appear promising as probes of heavy sector physics.  In the following
sections, we will therefore concentrate on measurements of the
super-oblique corrections through the observables described above,
that is, in processes involving gauginos.

\section{Uncertainties in Observables}
\label{sec:uncertainties}

Having now identified a large list of possible observables that depend
on the SUSY couplings $h_i$, we must determine if some of these may be
measured precisely enough to be significant probes of the heavy
sparticle sector.  In the sections that follow, we will consider such
quantitative issues in three examples that are representative in the
sense that there is one example for each coupling constant relation,
and one example for each of the three groups of particles given in
Sec.~\ref{subsec:epem}.  Here, however, we give a general description
of the various errors that enter such analyses and our treatment of
these errors.

The uncertainties may be grouped into categories.  First,
there are uncertainties arising from the many unknown SUSY parameters
that enter any given analysis.  These we will call theoretical
systematic uncertainties.  If, for example, a measurement of
super-oblique parameters is to be obtained from a cross section that
depends on $h_i$, the other parameters entering the cross section must
be carefully controlled.  These parameters include the masses of the
particles involved, as well as the field content of these particles,
for example, the gaugino content of relevant charginos and
neutralinos.  We will carefully study these errors, and will find
that, by appealing to other measurements and exploiting various
collider features, such uncertainties may be reduced to promisingly
low levels.

There are also uncertainties from finite experimental statistics and
backgrounds.  These will also be included, and we will present results
for specific integrated luminosities.  We assume that the backgrounds
are well-understood and so may be subtracted up to statistical
uncertainties.  This is a reasonable assumption for standard model
backgrounds.  Of course, for certain regions of parameter space, SUSY
backgrounds may enter.  These depend on {\em a priori} unknown SUSY
parameters, and the uncertainties associated with these are then part
of the first category discussed above.

In our analyses, we have not included radiative corrections in our
calculations of cross sections and branching ratios. The large
logarithm radiative corrections are absorbed in the super-oblique
parameters we are hoping to probe.  There remain, however, radiative
corrections from standard model particles, as well as the accessible
superpartners.  At the level of precision we will be considering,
these effects may be important.  However, these corrections are in
principle well-known once the calculations appropriate to the scenario
actually realized in nature are completed and a consistent one-loop
regularization scheme is established for all relevant observables.
Radiative corrections dependent on the light superparticles will be
subject to theoretical systematic uncertainties, but these are small
relative to the theoretical systematic uncertainties entering at tree
level, which were described above and will be included in our
analyses.

The final group of uncertainties are experimental systematic errors.
These include, for example, uncertainties in luminosity, detector
acceptances, initial state radiation effects, and, in some of the
measurements considered below, beam polarization and $b$-tagging
efficiency.  A complete analysis would require detailed experimental
simulations incorporating all of these experimental systematic
uncertainties.  Such an analysis is beyond the scope of this work,
especially since the sizes of some of these uncertainties at the NLC
are unknown and are currently under investigation.  We will see,
however, that in some cases the experimental systematic uncertainties
are likely to be negligible relative to the errors described above;
where this is not the case, we will note which experimental systematic
errors appear to be most important.  By estimating the sizes of the
errors from the sources described in the paragraphs above, we will
find interesting implications for what collider specifications are
required and what features are particularly promising for the study of
non-decoupling SUSY breaking effects.

\section{Probe of SU(2) Couplings from charginos}
\label{sec:charginos}

In this section, we consider a probe of the SU(2) relation $g_2 =
h_2$.  Recall from Sec.~\ref{sec:introduction} that the size of
deviations from this equivalence may be parametrized by the
super-oblique parameter $\susyU_2$, which, in the two scenarios we
considered, is

\begin{equation}
\susyU_2 \equiv h_2/g_2 - 1 \approx 0.7-0.8\% \times \ln \frac{M}{m} 
\ .
\end{equation}
For a light sector scale $m \approx {\cal O}(100 \text{ GeV})$, we see
that measurements of $\susyU_2$ to accuracies of 3--4\% are required
to be sensitive to deviations from a heavy scale $M \approx {\cal O}
(10 \text{ TeV})$, while determination of the heavy scale to within a
factor of 3 requires measurements at the 0.8--0.9\% level.  Of course,
larger deviations from greater $M$ or additional exotic
supermultiplets are possible, but we will take these figures as useful
reference points.

As a test of the SU(2) coupling relation, we turn to the first group
of sparticles given in Sec.~\ref{sec:observables}, charginos and
neutralinos, and consider chargino pair production at the NLC.  This
process is promising, as charginos are typically among the lighter
sparticles, and they are produced with large cross section when
kinematically accessible.  In addition, in our scenarios, the
constraint of the $Z$ mass implies that the Higgsino mass parameter
$|\mu|$ is usually of order the third generation squark masses.  This
often implies that the lighter chargino and neutralinos are
gaugino-like, and is exactly the region of parameter space where we
have some hope of measuring $h_2$ accurately with charginos, as
explained in Sec.~\ref{sec:observables}.

The measurement of $h_2$ from chargino production was previously
considered in Ref.~\cite{FMPT}, and we therefore begin with a review
of those results.  Details, particularly those concerning the error
analysis, will be omitted, and we refer interested readers to the
original study for a complete treatment.  In Ref.~\cite{FMPT}, the
following parameters were taken as a case study in the gaugino region:

\begin{equation}
(\mu, M_2, \tan\beta, M_1/M_2, m_{\sneutrino_e}) 
= (-500 \text{ GeV}, 170 \text{ GeV}, 4, 0.5, 400 \text{ GeV}) \ .
\end{equation}
With these parameters, the light chargino and neutralino masses are
$m_{\charginoone} = 172 \text{ GeV}$ and $m_{\LSP} = 86 \text{ GeV}$,
and the cross sections for chargino pair production with $\sqrt{s} =
500 \text{ GeV}$, unpolarized $e^+$ beams, and right- and
left-polarized $e^-$ beams are $\sigma_R = 0.15 \text{ fb}$ and
$\sigma_L = 612 \text{ fb}$.  As is characteristic of the gaugino
region, $\sigma_R$ is highly suppressed, but $\sigma_L$ is large.
With design luminosity ${\cal L} = 50 \text{ fb}^{-1}/\text{yr}$, tens
of thousands of charginos will be produced each year, giving us hope
that ${\cal O}(1)\%$ measurements may be feasible.  Finally, the decay
$\charginoone \to W^{\pm} \LSP$ is open and dominant --- the chargino
branching ratios are therefore equivalent to those of the
$W$.\footnote{For extremely large values of $|\mu|$, the chargino and
neutralino are nearly pure gauginos, and the on-shell $W$ decay mode
may be so suppressed that decays through off-shell sleptons and
squarks significantly shift the chargino branching ratios. We will not
consider this case, but note that such a scenario typically requires
values of $|\mu|$ far above the TeV scale and would itself be a
striking signature for heavy mass scales.}

Charginos may be produced through $t$-channel sneutrino exchange and
$s$-channel $\gamma$ and $Z$ diagrams.  The first amplitude depends on
$h_2^2$, and is the source of our sensitivity to super-oblique
corrections.  The left-polarized differential cross section is
therefore dependent on 5 parameters beyond the standard model:

\begin{equation}
\frac{d\sigma}{d\cos \theta} \left(e^-_L e^+ \to \tilde{\chi}_1^+
\tilde{\chi}_1^- \right) =
\frac{d\sigma}{d\cos \theta} \left( m_{\chargino}, \phi_+, \phi_-,
m_{\sneutrino_e}, h_2 \right) \ ,
\end{equation}
where the angles $\phi_{\pm}$ specify the composition of
$\charginoone$ in terms of the weak eigenstates $(-i\tilde{W}^{\pm},
\tilde{H}^{\pm})$.  To measure $h_2$, we must also constrain the other
parameters.  The mass $m_{\charginoone}$ may be measured to 2 GeV by
determining energy distribution endpoints of the decay
products~\cite{JLC}.  The Wino-ness of the chargino may be established
by checking that $\sigma_R \approx 0$.  Alternatively, one can verify
that $\tilde{\chi}^{\pm}_1\tilde{\chi}^{\mp}_2$ production is
kinematically inaccessible, which puts lower limits on $|\mu|$ and the
gaugino-ness of the chargino.  (Of course, if higher beam energy is
available, one could discover the heavy chargino or neutralinos and
measure $|\mu|$ and the angles $\phi^{\pm}$.)  The resulting errors in
$m_{\charginoone}$ and $\phi_{\pm}$ at a $\sqrt{s}=500
\text{ GeV}$ machine were studied in Ref.~\cite{FMPT} and were 
found to be negligible relative to the uncertainties we now describe.

\begin{figure}
\centerline{\psfig{file=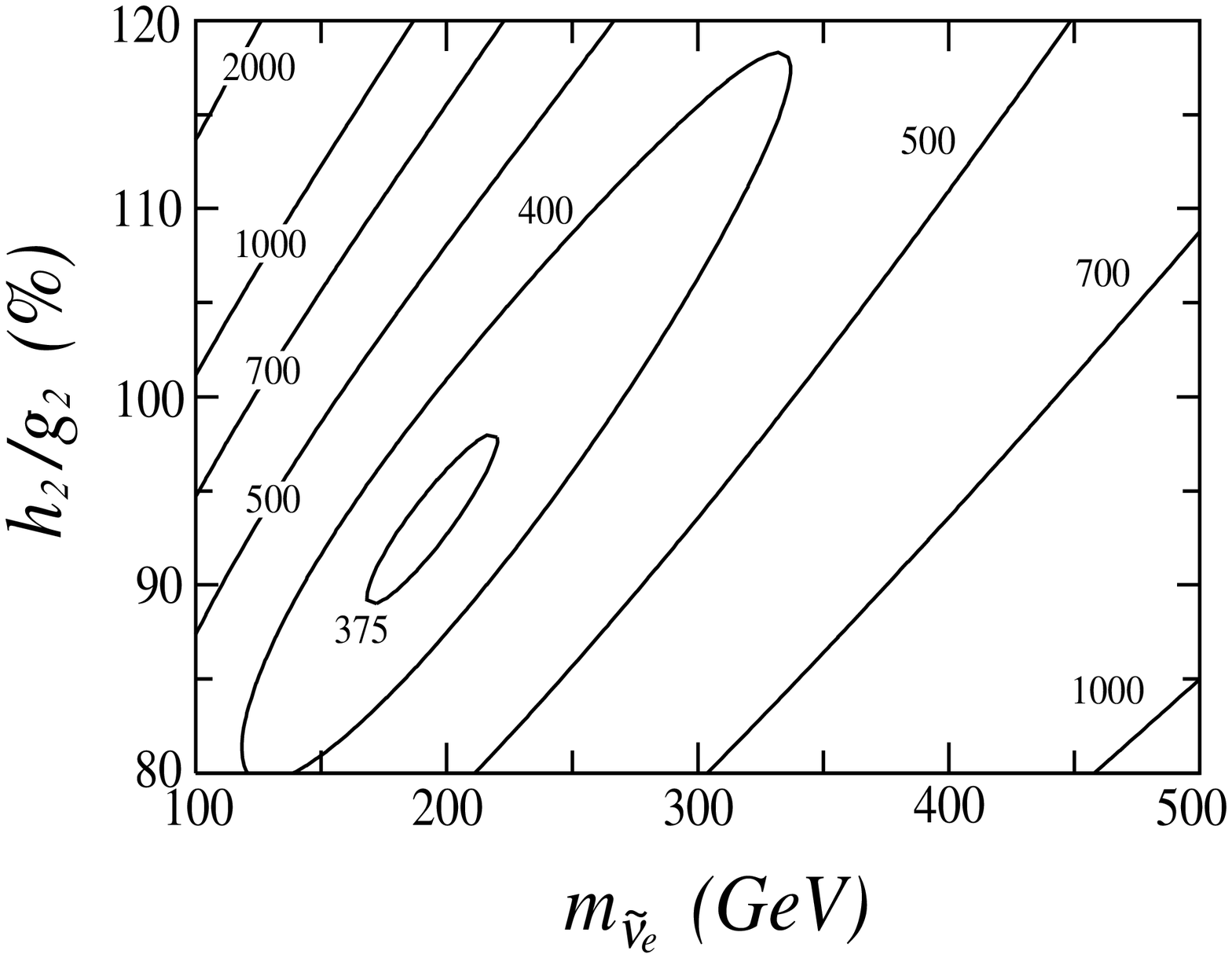,width=0.7\textwidth}}
\vspace*{.15in}
\caption{Contours of constant chargino pair production cross section 
$\sigma_L$ in fb in the $(m_{\sneutrino_e}, h_2)$ plane for underlying
parameters $(\mu, M_2, \tan\beta, M_1/M_2) = (-500 \text{ GeV}, 170
\text{ GeV}, 4, 0.5)$ and $\protect\sqrt{s}=500\text{ GeV}$.
\label{fig:sigma500}}
\end{figure}

\begin{figure}
\centerline{\psfig{file=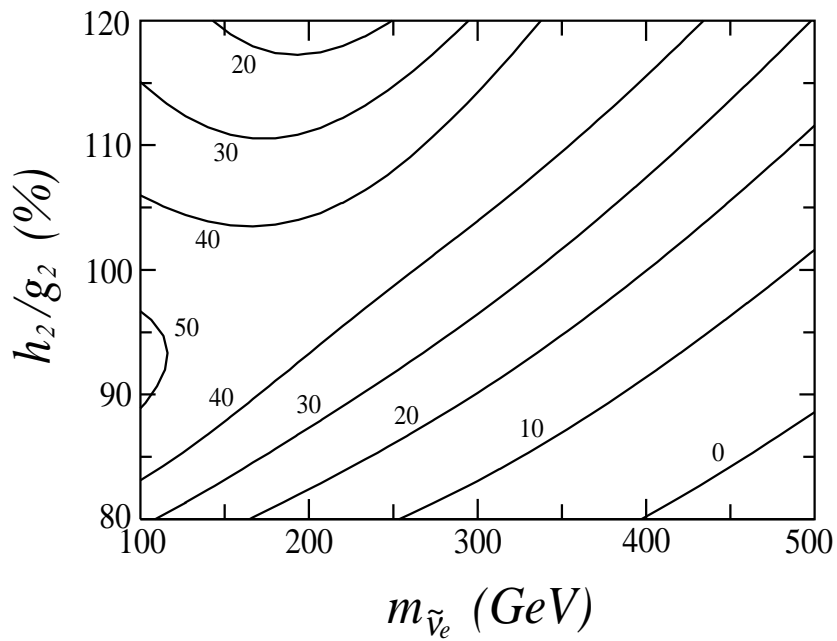,width=0.7\textwidth}}
\vspace*{.15in}
\caption{Contours of constant chargino forward-backward asymmetry 
$A_L^{\chi}$ in percent in the $(m_{\sneutrino_e}, h_2)$ plane for
underlying parameters $(\mu, M_2, \tan\beta, M_1/M_2) = (-500 \text{
GeV}, 170 \text{ GeV}, 4, 0.5)$ and $\protect\sqrt{s}=500\text{ GeV}$.
\label{fig:afb500}}
\end{figure}

The remaining two unknowns, $m_{\sneutrino_e}$ and $h_2$, may then be
determined by the $e^-_L$ total cross section $\sigma_L$ and a
truncated forward-backward asymmetry

\begin{equation}
A_L^{\chi} = \frac{\sigma_L (0<\cos \theta < 0.707) -
\sigma_L(-1<\cos\theta < 0)}{\sigma_L(-1<\cos\theta<0.707)} \ .
\end{equation}
This peculiar definition of $A_L^{\chi}$ is dictated by cuts designed
to remove the forward-peaked $W$ pair production. These two quantities
are plotted in Figs.~\ref{fig:sigma500} and \ref{fig:afb500} for
$\sqrt{s} = 500\text{ GeV}$.  Unfortunately, these quantities cannot
be measured directly.  To determine them, we look at mixed mode
events, where one chargino decays hadronically and the other
leptonically.  $A_L^{\chi}$ is measured through its correlation with
the observed forward-backward asymmetry of the hadronic decay products
$A^{\text{had}}$, and the total cross section is determined by its
correlation with the measured mixed mode cross section after cuts.
Both of these correlations are imperfect.  The correlation between
$A^{\text{had}}$ and $A_L^{\chi}$ has a slight dependence on
additional SUSY parameters entering the decay, such as $M_1$.  The
total cross section determination is weakened by its dependence on the
cut efficiencies, which also depend on these additional SUSY
parameters.\footnote{Note that the determination of the total cross
section from the mixed cross section also requires that the chargino
branching fractions be known.  If decays through on-shell $W$ bosons
are closed, the branching ratios must also be determined by
considering the purely hadronic or purely leptonic modes, introducing
additional uncertainties that may significantly weaken the results.}
These effects lead to theoretical systematic errors, which are
investigated by Monte-Carlo simulations, where the lack of correlation
is determined by varying all the relevant SUSY parameters throughout
their ranges, subject only to the constraint that they reproduce
various observables, such as the chargino mass, within the
experimental errors.

In addition to these theoretical systematic errors, uncertainties from
backgrounds, dominated by $WW$ production, and finite statistics must
be included.  The resulting 1$\sigma$ uncertainties are~\cite{FMPT}

\begin{eqnarray}
\Delta A^{\chi}_L &=& 0.067\, (0.048)\, [0.037] \nonumber \\
\frac{\Delta \sigma_L}{\sigma_L} &=& 7.2\, (5.6)\, [4.7] \% \ ,
\label{charginobounds1} 
\end{eqnarray}
where the first two uncertainties are for integrated luminosities of
30 (100) $\text{fb}^{-1}$, and the final bracketed uncertainties are
from systematic errors alone, {\em i.e.}, the uncertainties in the
limit of infinite statistics.  Given these values, the expected $\sim
1\%$ uncertainty in luminosity~\cite{JLC} is negligible.  If similar
uncertainties in beam polarization may be obtained, they too have
little impact.  In any case, note that beam polarization is used here
only to increase the effective luminosity for this study, as the
signal and leading $WW$ background both exist only for left-polarized
beams.  Thus, if polarization uncertainties are dominant, the systematic
error from this source may be eliminated by using an unpolarized
beam, with a resultant decrease in effective luminosity by a factor of
2.

\begin{figure}
\centerline{\psfig{file=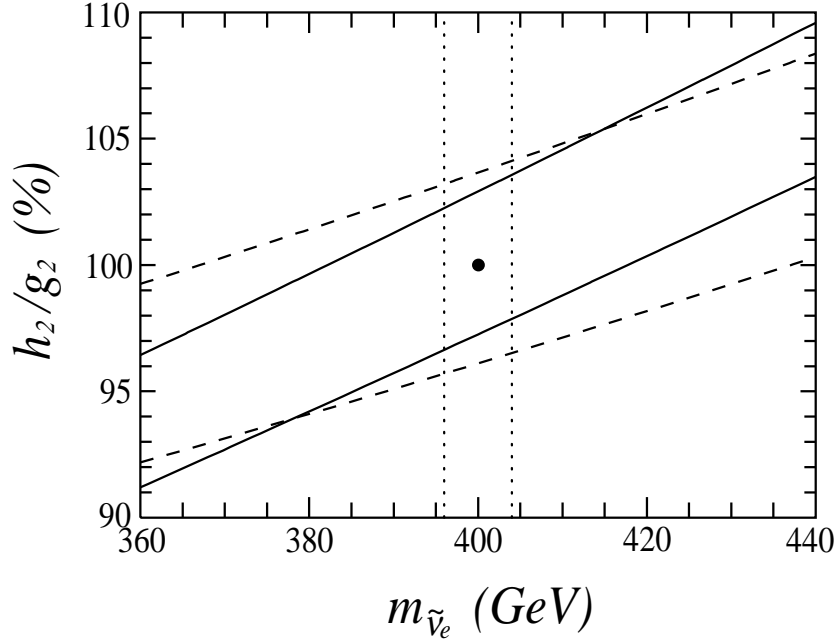,width=0.7\textwidth}}
\vspace*{.15in}
\caption{The allowed region in the $(m_{\sneutrino_e}, h_2)$ plane for
$\protect\sqrt{s}=500\text{ GeV}$ and $L = 100\text{ fb}^{-1}$. The
solid (dashed) curves are 1$\sigma$ contours of constant $\sigma_L$
($A^{\chi}_L$), and the underlying parameter point $(\mu, M_2,
\tan\beta, M_1/M_2, m_{\sneutrino_e}) = (-500 \text{ GeV}, 170 \text{
GeV}, 4, 0.5, 400 \text{ GeV})$ is indicated.  The allowed region is
bounded by the $\sigma_L$ and $A^{\chi}_L$ contours and the bound on
$m_{\sneutrino_e}$; for reference, the bound $\Delta m_{\sneutrino_e}
= 4\text{ GeV}$ is given by the dotted contours.
\label{fig:chi500400}}
\end{figure}

The measurements of Eq.~(\ref{charginobounds1}) determine allowed
regions in the $(m_{\sneutrino_e}, h_2)$ plane, which we define
crudely to be regions that are within the 1$\sigma$ contours of all
observables.  The relevant region for integrated luminosity $100
\text{ fb}^{-1}$ is given in Fig.~\ref{fig:chi500400}.  Even without a
measurement of $m_{\sneutrino_e}$, we see that the ratio $h_2/g_2$ is
constrained to be consistent with unity, a quantitative confirmation
of SUSY and the interpretation that the fermion being studied is in
fact the chargino.

The measurements at $\sqrt{s}=500\text{ GeV}$ also bound the
sneutrino's mass through its virtual effects.  With this strong
motivation, one would then increase the beam energy to find
$\sneutrino_e$ pair production. Studies have found that $\sim 1\%$
measurements of charged slepton masses are possible at the
NLC~\cite{BV}, and similar levels have been achieved in sneutrino
studies through measurements of electron energies in the decay
$\sneutrino_e \to e^{\mp}\charginoone$~\cite{SnowmassNLC}.  With this
as an additional constraint, we may return to Fig.~\ref{fig:chi500400}
and look for small deviations from $g_2=h_2$.  We see that, for
example, if $m_{\sneutrino_e}$ is measured to 4 GeV, deviations of
$\susyU_2$ from its central value are constrained to the range

\begin{equation}
-3\% < \Delta \susyU_2 < 3\% \quad (m_{\sneutrino_e} = 400 
\text{ GeV}, \sqrt{s} = 500 \text{ GeV}) \ .
\end{equation}
At this parameter point, the determination is sufficiently accurate
that to good approximation, the uncertainties are linear, {\em i.e.},
if the underlying value of $\susyU_2$ is 4\%, the allowed range is
$1\% < \susyU_2 < 7\%$.  Thus, if the mass of squarks is $\agt {\cal
O} (10\text{ TeV})$, deviations from exact SUSY may be seen and
$\susyU_2$ may be bounded to be positive.  Such a measurement would
provide unambiguous evidence for very massive superparticle
states. Note, however, that the mass scale of such states is
determined only to a couple of orders of magnitude.

\begin{figure}
\centerline{\psfig{file=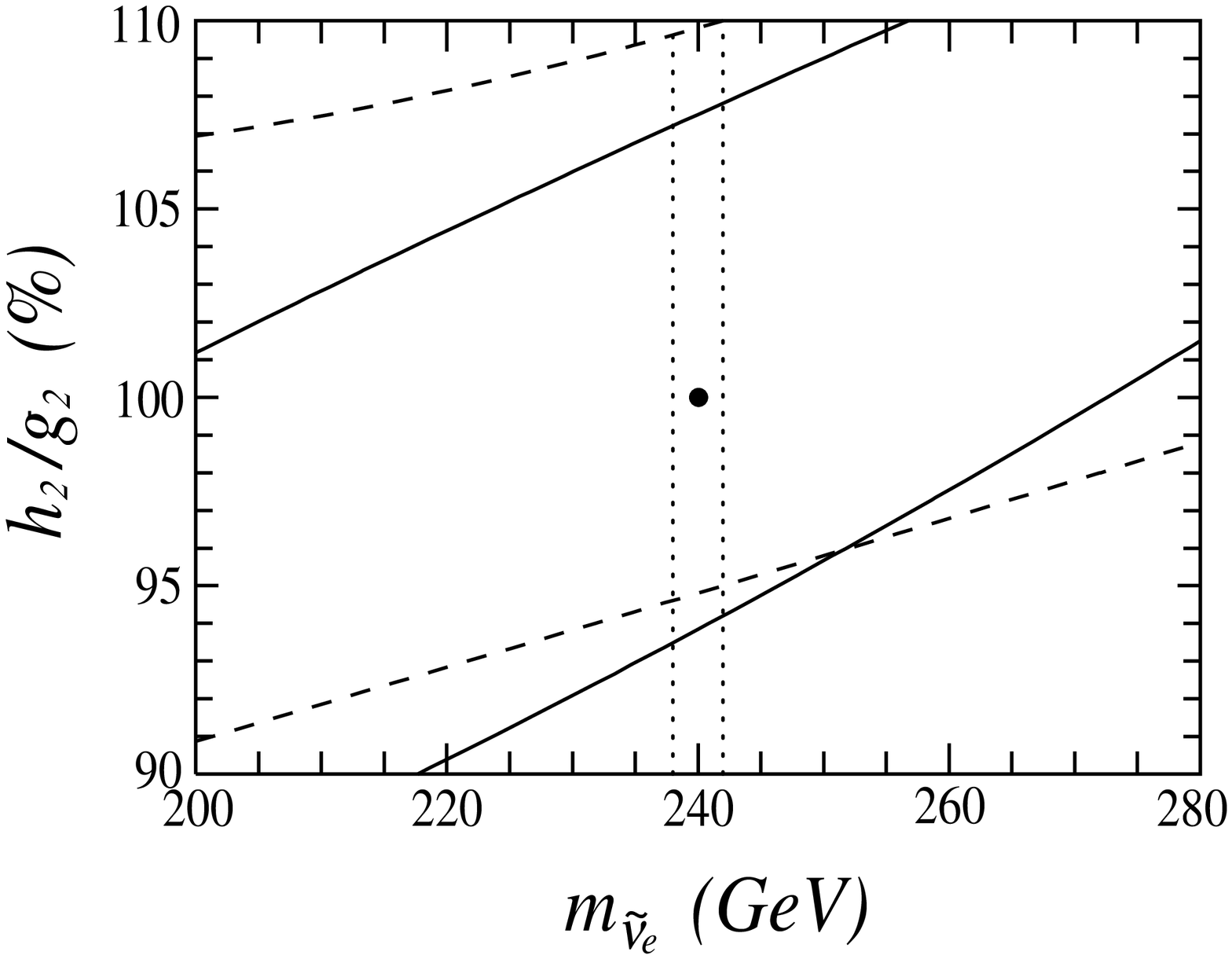,width=0.7\textwidth}}
\vspace*{.15in}
\caption{The same as in Fig.~\protect\ref{fig:chi500400}, but with 
underlying parameter $m_{\sneutrino_e} = 240 \text{ GeV}$, and
dotted contours at $\Delta m_{\sneutrino_e} = 2\text{ GeV}$.
\label{fig:chi500240}}
\end{figure}

In fact, the precision of the above study may be improved by
exploiting an important feature of the NLC, its adjustable beam
energy.  To illustrate this most vividly, let us consider another point
in parameter space with a different sneutrino mass.  In
Ref.~\cite{FMPT}, a large $m_{\sneutrino_e}$ was chosen to illustrate
the sensitivity of precision measurements to effects of virtual
sparticles.  For $\sqrt{s} = 500\text{ GeV}$ and $m_{\sneutrino_e} =
400\text{ GeV}$, $\sigma_L$ and $A_L^{\chi}$ are quite sensitive to
changes in $h_2$ and $m_{\sneutrino_e}$.  However, for other
underlying parameters, this may not be the case.  For example, we see
in Figs.~\ref{fig:sigma500} and \ref{fig:afb500} that, for $\sqrt{s} =
500 \text{ GeV}$ and $m_{\sneutrino_e} = 240\text{ GeV}$, $\sigma_L$
is near a minimum and $A_L^{\chi}$ is near a saddle point at
$h_2=g_2$. Thus for such a sneutrino mass, there are relatively few
events, and more importantly, the dependence of our observables on
$h_2$ is weak.  By carrying out the analysis outlined above for this
new parameter point, we find

\begin{eqnarray}
\Delta A^{\chi}_L &=& 0.079\, (0.053) \nonumber \\
\frac{\Delta \sigma_L}{\sigma_L} &=& 9.4\, (6.2) \% \ ,
\end{eqnarray}
where these 1$\sigma$ uncertainties are for integrated luminosities of
30 (100) $\text{fb}^{-1}$.  (We have assumed here that the theoretical
systematic errors in this case are as in the previous
$m_{\sneutrino_e} = 400\text{ GeV}$ analysis.  This assumption is
valid, as these uncertainties are not dominant, and are in any case
most sensitive to quantities, such as the chargino velocity, that are
identical in these two case studies.)  In Fig.~\ref{fig:chi500240}, we
plot the region allowed by these measurements.  The determination of
$\susyU_2$ is greatly deteriorated.  If $m_{\sneutrino_e}$ is again
measured to $\sim 1\%$, the range of $\susyU_2$ in the allowed region is
(taking a central value of $\susyU_2=0$)

\begin{equation}
-5\% < \susyU_2 < 8\% \quad (m_{\sneutrino_e} = 240 \text{ GeV},
\sqrt{s} = 500 \text{ GeV}) \ .
\end{equation}

\begin{figure}
\centerline{\psfig{file=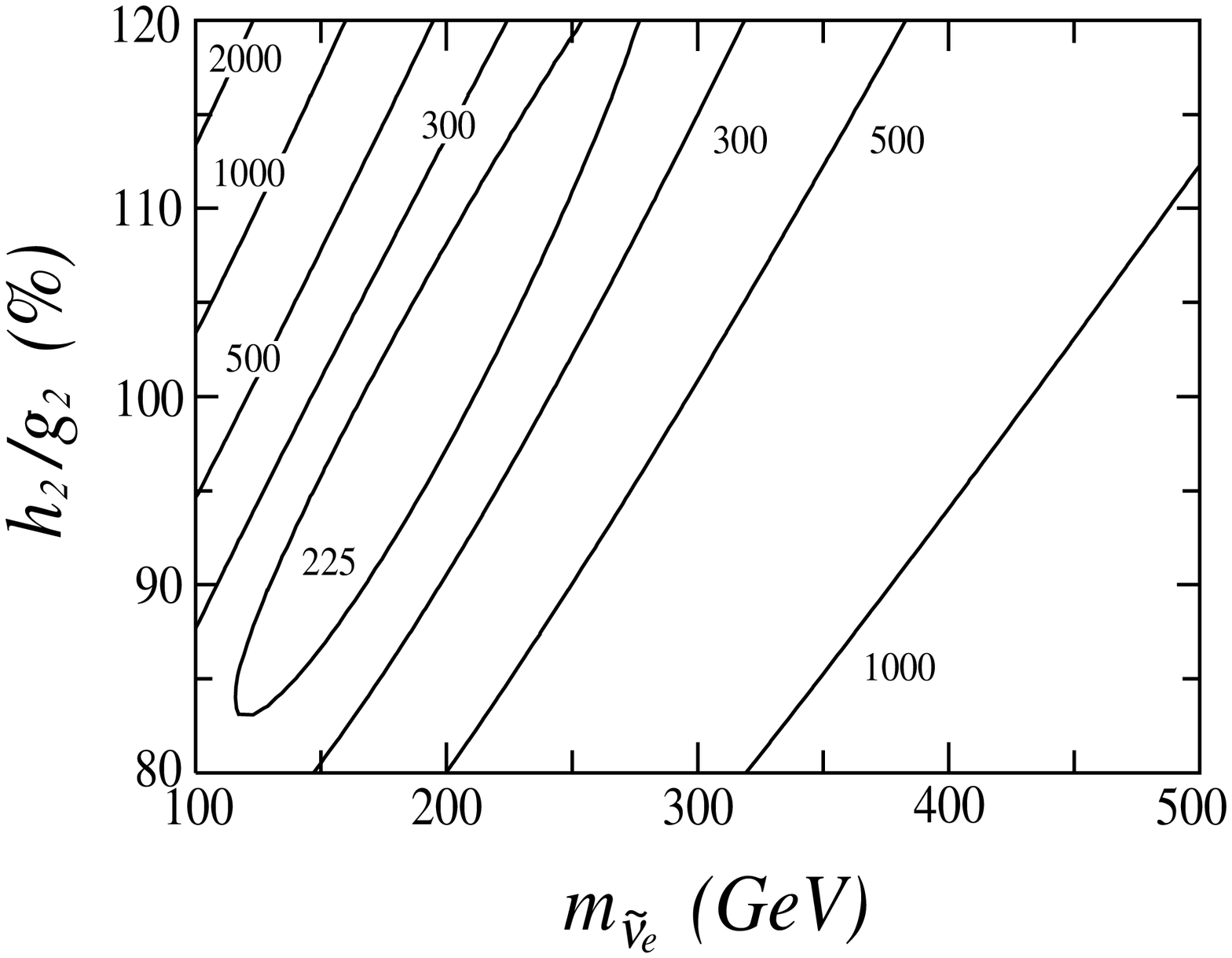,width=0.7\textwidth}}
\vspace*{.15in}
\caption{Same as in Fig.~\protect\ref{fig:sigma500}, but for 
$\protect\sqrt{s}=400\text{ GeV}$.
\label{fig:sigma400}}
\end{figure}

\begin{figure}
\centerline{\psfig{file=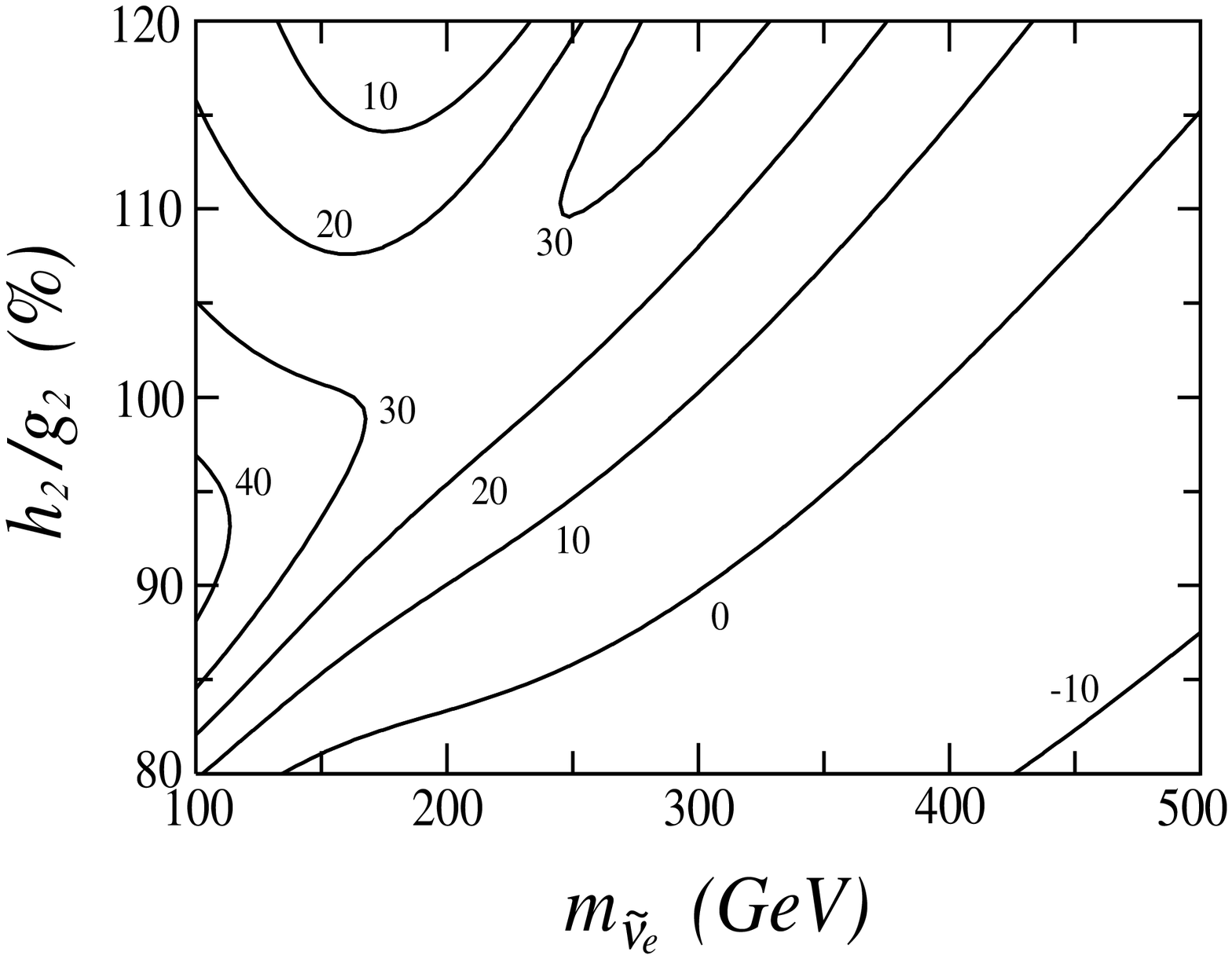,width=0.7\textwidth}}
\vspace*{.15in}
\caption{Same as in Fig.~\protect\ref{fig:afb500}, but for 
$\protect\sqrt{s}=400\text{ GeV}$.
\label{fig:afb400}}
\end{figure}

\begin{figure}
\centerline{\psfig{file=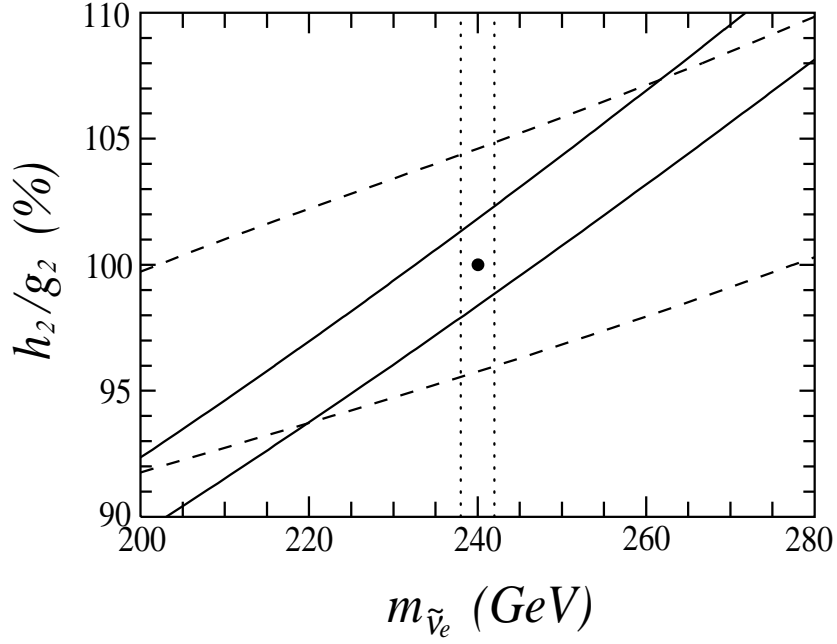,width=0.7\textwidth}}
\vspace*{.15in}
\caption{The same as in Fig.~\protect\ref{fig:chi500400}, but with 
underlying parameter $m_{\sneutrino_e} = 240 \text{ GeV}$, dotted
contours at $\Delta m_{\sneutrino_e} = 2\text{ GeV}$, and improved
center-of-mass energy $\protect\sqrt{s}= 400 \text{ GeV}$.
\label{fig:chi400240}}
\end{figure}

The underlying SUSY parameters above appear to lead to poor bounds on
super-oblique corrections.  However, an important aspect of $\epem$
colliders is the ability to adjust the initial state parton energy.
This flexibility may be used to eliminate backgrounds, and also to
improve the sensitivity to underlying parameters.  Here, we exploit
the latter virtue.  The extrema in $\sigma_L$ and $A_L^{\chi}$ may be
shifted by choosing different beam energies.  In
Figs.~\ref{fig:sigma400} and
\ref{fig:afb400}, we plot $\sigma_L$ and $A_L^{\chi}$ in the
$(m_{\sneutrino_e}, h_2)$ plane again, but now for $\sqrt{s} = 400
\text{ GeV}$. We see that the extrema in the $\sigma_L$ and
$A^{\chi}_L$ observables are shifted to lower $m_{\sneutrino_e}$, and
the strong dependence of $\sigma_L$ and $A_L^{\chi}$ on $h_2$ for
$m_{\sneutrino_e} = 240\text{ GeV}$ is restored.  Applying the same
analysis once again, we find, including all theoretical systematic and
experimental statistical errors,

\begin{eqnarray}
\Delta A^{\chi}_L &=& 0.11\, (0.068) \nonumber \\
\frac{\Delta \sigma_L}{\sigma_L} &=& 11\, (7.3) \% \ ,
\end{eqnarray}
for integrated luminosities of 30 (100) $\text{fb}^{-1}$.\footnote{In
arriving at these results, we have not designed optimized cuts for
$\sqrt{s} = 400\text{ GeV}$, but have simply assumed that the
efficiency of the cuts for the $WW$ background is unchanged at
$\sqrt{s}=400\text{ GeV}$.  The results are rather insensitive to this
assumption; for example, making the highly pessimistic assumption that
the background is in fact doubled leads to $\Delta A^{\chi}_L = 0.083$
and $\frac{\Delta \sigma_L}{\sigma_L} = 8.9 \%$ for 100 fb$^{-1}$.}
We see that these uncertainties are larger than at $\sqrt{s} = 500
\text{ GeV}$.  However, the increased sensitivity of $\sigma_L$ and 
$A^{\chi}_L$ to $h_2/g_2$ more than makes up for the loss in
statistics, as can be seen in Fig.~\ref{fig:chi400240}, where we plot
the allowed region for underlying parameters as in
Fig.~\ref{fig:chi500240}, but for $\sqrt{s} = 400 \text{ GeV}$.
Assuming again a $\sim 1\%$ measurement of $m_{\sneutrino_e}$, the
range of allowed deviations of $\susyU_2$ from its central value in
the allowed region is

\begin{equation}
-2\% < \Delta \susyU_2 < 2\% \quad (m_{\sneutrino_e} = 240 
\text{ GeV}, \sqrt{s} = 400 \text{ GeV}) \ ,
\end{equation}
where again we have checked that the uncertainties are linear.  Such a
measurement gives one an extremely precise measurement of $h_2$, and
even begins to provide interesting constraints on the heavy squark
scale for the purposes of model-building. Note that this bound from
charginos is comparable to the previous bound derived from selectron
production in the $\epem$ mode of linear colliders~\cite{slepton}.
The bound from selectron production was $\sim 1\%$ on the parameter
$\susyU_1$, which we expect in typical models to be roughly half as
sensitive to the effects of heavy superpartners.

Although a complete scan of parameter space is beyond the scope of
this study, we see that if gaugino-like charginos are produced at the
NLC, interesting bounds on the super-oblique parameter $\susyU_2$ may
be obtained.  Such bounds rely on a variety of precise measurements
constraining the gaugino content of the chargino and the
$\sneutrino_e$ mass.  In addition, we have seen that the sensitivity
of observables to the super-oblique parameters may be markedly
improved by adjusting the beam energy.  Given a better understanding
of the uncertainties obtainable in the sneutrino mass and various
experimental systematic uncertainties, the beam energy may be
optimized to increase the sensitivity to super-oblique corrections and
multi-TeV superpartners.

\section{Probe of U(1) Couplings from selectrons}
\label{sec:selectrons}

In this section, we consider measurements of the U(1) gaugino coupling
$h_1$ from selectron production. From Sec.~\ref{sec:introduction}, we
see that the deviation between the U(1) gauge boson and gaugino couplings
for the heavy QCD and 2--1 models is

\begin{equation}
\susyU_1 \equiv h_1/g_1 - 1 \approx 0.3-0.35\% \times \ln \frac{M}{m}
\ .
\end{equation}
For a heavy scale in the multi-TeV range, the deviation is about 1\%.
A determination of the heavy scale to within a factor of 3 requires
the precision of the $\susyU_1$ measurement to be at the $\sim 0.3\% $
level, which will be taken as our target precision.  The effects are
clearly smaller than in the SU(2) and SU(3) cases and require
correspondingly more precise measurements for similar bounds on the
heavy mass scale.

The possibility of measuring $h_1$ from $\selectron_R$ production in
$\epem$ collisions at a linear collider has been considered previously
in Ref.~\cite{slepton}, where bounds from
the differential cross section $d\sigma (\epem \to \selectron^+_R
\selectron^-_R)/ d \cos\theta$ were found to imply bounds on
$\susyU_1$ at the $\sim 1\%$ level.  As was pointed out in
Ref.~\cite{slepton}, such a measurement provides an extremely high
precision test of SUSY, and may possibly provide evidence for
decoupling effects from heavy sectors.  However, as the expected
super-oblique corrections in the U(1) sector are small, such a test,
as in the chargino case considered in the previous section, is
probably not sufficient to determine the heavy superpartner scale to
better than an order of magnitude.

\begin{figure}
\centerline{\psfig{file=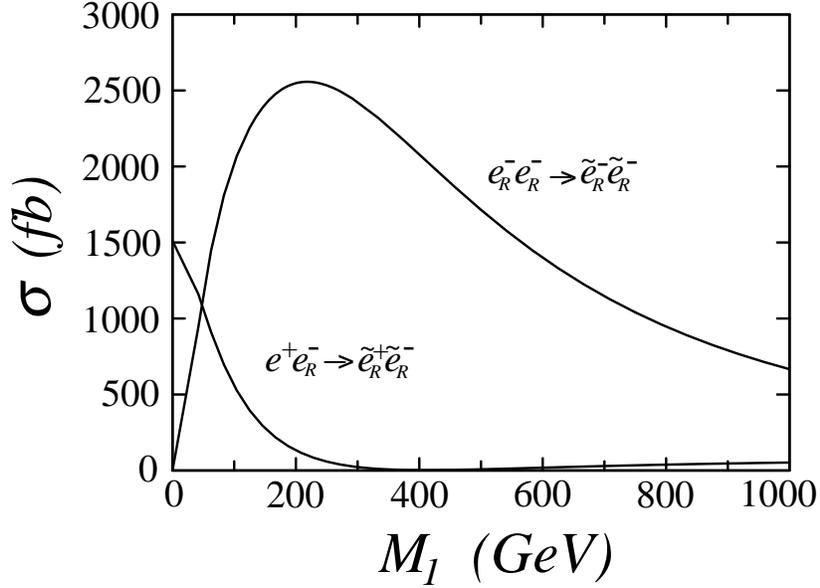,width=0.7\textwidth}}
\vspace*{.15in}
\caption{The total selectron pair production cross sections for the
$e_R^- e_R^-$ and $e^+ e_R^-$ modes with $m_{\selectron_R}=150$ GeV
and $\protect\sqrt{s}=500$ GeV, as functions of the Bino mass $M_1$,
assuming the Bino is a mass eigenstate. Note that the very small (but
nonzero) cross section for the $e^+ e_R^-$ mode near $M_1 \sim 400$
GeV results from destructive interference between the $s$- and
$t$-channel diagrams.
\label{fig:M1dependence}}
\end{figure}

To increase this sensitivity, we consider here $\selectron_R$ pair
production in the $\emem$ mode of a future linear collider.  (The
extension to $\selectron_L$ is straightforward and will be discussed
at the end of this section.) There are several advantages in
considering selectron production at an $\emem$ collider:
\begin{itemize}

\item At an $\emem$ collider, selectrons are produced only through 
$t$-channel neutralino exchange. The cross section for $\selectron_R$
production is thus directly proportional to $h_1^4$. In contrast, at
$\epem$ colliders, selectrons are produced through both $s$- and
$t$-channel processes.  The $s$-channel processes are $h_1$
independent, and may significantly dilute the sensitivity of the cross
section observables to variations in $h_1$.

\item The backgrounds to selectron pair production at $\emem$
colliders are very small.  Most of the major backgrounds present in
the $\epem$ mode are absent; {\it e.g.}, $W$ pair and chargino pair
production are forbidden by total lepton number conservation. This
makes the $\emem$ environment extremely clean for precision
measurements.

\item It is possible to highly polarize both $e^-$ beams.  
Polarizing both beams right-handed increases the desired $\selectron_R
\selectron_R$ cross section by a factor of 4, and suppresses remaining
backgrounds, such as $e^- \nu W^-$, even further.

\item In order to produce $\selectron_R^- \selectron_R^-$, a Majorana 
mass insertion in the neutralino propagator is needed to flip the
chirality.  The total cross section therefore increases as the Bino
mass $M_1$ increases as long as $M_1$ is not too large ($M_1 \alt
\sqrt{s}/2$).  The $M_1$ dependences of the cross sections for $e^+
e_R^- \to \selectron_R^+ \selectron_R^-$ and $e_R^- e_R^- \to
\selectron_R^- \selectron_R^-$ are shown in 
Fig.~\ref{fig:M1dependence} for $\sqrt{s} = 500 \text{ GeV}$ and
$m_{\selectron_R} = 150 \text{ GeV}$. One can see that if $M_1$ is not
too small, the selectron production cross section in the $\emem$ mode
is much larger than in the $\epem$ mode.\footnote{It is interesting to
note that this dependence may allow an alternative high mass scale
probe in the Higgsino region where $|\mu| < M_1, M_2$ and gaugino
masses may be very large.  If selectron pairs may be produced, their
pair-production cross section in the $\emem$ mode is still substantial
and sensitive to $M_1$ even for very large $M_1$, and may be used to
determine values of $M_1$ at the multi-TeV scale. Here, however, we
assume that we are in the gaugino region since we are interested in
measuring the gaugino couplings.}  This compensates for any reduction
in luminosity that may be present in the $\emem$ mode.

\item The $t$-channel gaugino mass insertion may also be exploited to
reduce theoretical systematic errors arising from uncertainties in the
$\selectron_R$ and $\LSP$ masses. The $\selectron_R$ and $\LSP$ masses
are typically constrained from electron energy distribution endpoints.
The resulting allowed masses are positively correlated, while the
dependence of the total cross section in the $\emem$ mode on
$m_{\LSP}$ and $m_{\selectron_R}$ is negatively correlated (in the
region of the parameter space in which we are interested). The total
cross section may therefore remain approximately constant over the
allowed region in the $(m_{\selectron_R}, m_{\LSP})$ plane. This point
will be described in more detail below.
\end{itemize}

Let us now consider quantitatively the possibility of precisely
measuring $h_1$ using an $\emem$ collider. We will determine $h_1$
from the total cross section $\sigma_R = \sigma(e_R^- e_R^- \to
\selectron_R^- \selectron_R^-)$.  We assume that the $\selectron_R$ 
decays directly to $e\LSP$, and that $\LSP$ is the lightest
supersymmetric particle and is Bino-like. The cross section is
proportional to $h_1^4$, so in order to measure $h_1$ to 0.3\%, the
cross section must be determined to 1.2\%. There are many possible
sources of uncertainties, as was mentioned in
Sec.~\ref{sec:uncertainties}. The experimental statistical and
systematic errors will introduce uncertainties in determining
$\sigma_R$ experimentally. Once $\sigma_R$ is determined, the
extraction of $h_1$ from this measurement depends on many other
unknown SUSY parameters and hence suffers from theoretical systematic
uncertainties.  To achieve the target precision, each source of
uncertainty should induce an error in $\sigma_R$ less than 1\%.  Of
course, if there are several comparable uncertainties, they are
required to be even smaller so that their combined error is at the 1\%
level.

\begin{figure}
\centerline{\psfig{file=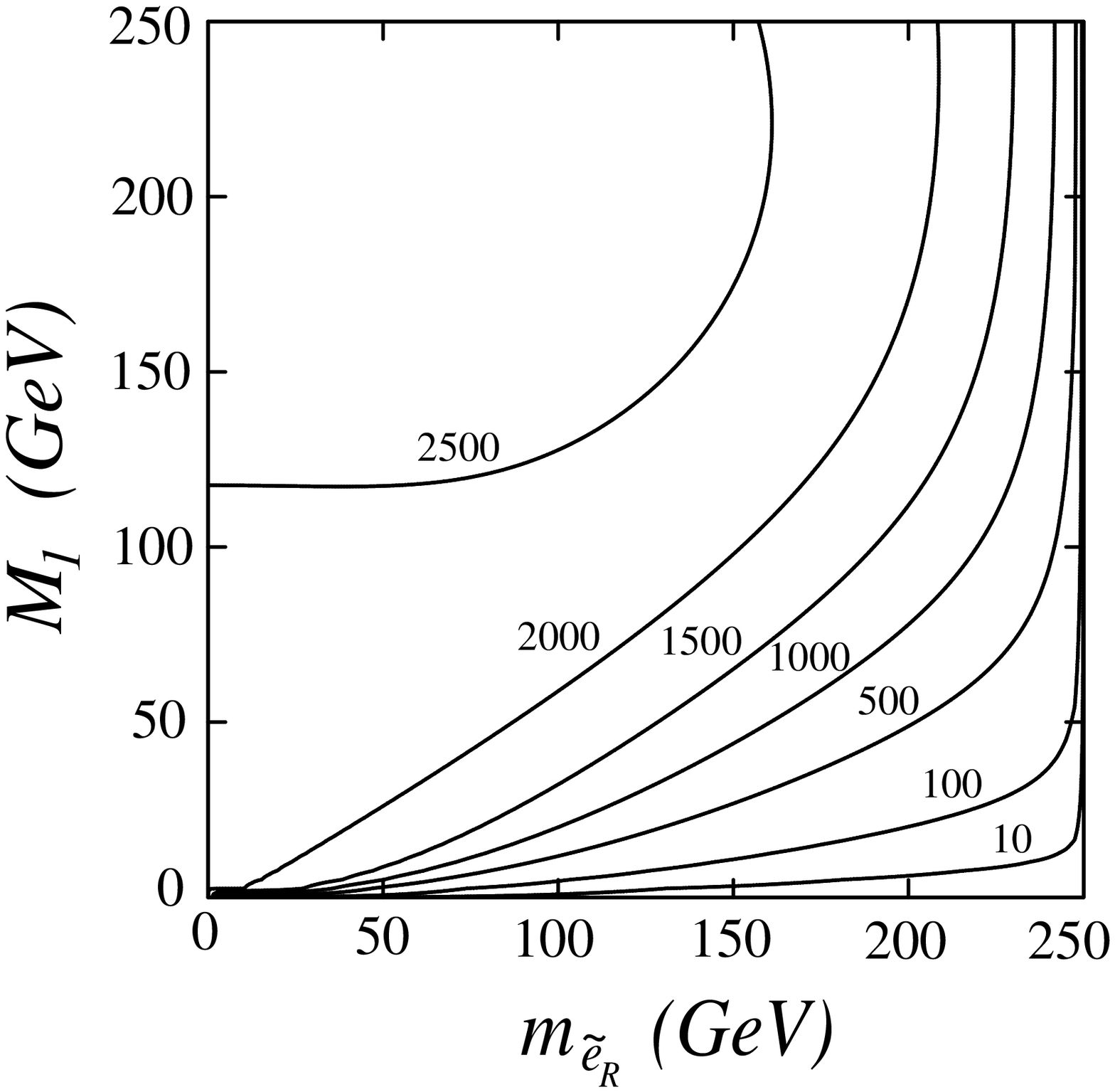,width=0.6\textwidth}}
\vspace*{.15in}
\caption{Contours of constant $\sigma_R = \sigma (e^-_R e^-_R \to 
\selectron^-_R \selectron^-_R)$ in fb in the ($m_{\selectron_R}$, 
$M_1$) plane for $\protect\sqrt{s}=500$ GeV.
\label{fig:emxsec}}
\end{figure}

The possible sources of uncertainties in measuring $\sigma_R$ include:

1. Statistical fluctuation: Fig.~\ref{fig:emxsec} shows the total
cross section $\sigma_R$ in the $(m_{\selectron_R}, M_1)$ plane. We
can see that for a significant part of the parameter space ($M_1$ not
too small and $m_{\selectron_R}$ not too close to threshold), the
total cross section is on the order of $\sim 2000$ fb.  Typically only a
small fraction of the selectrons are produced along the beam direction
($< 5\%$ for $\sin \theta (\selectron_R)< 5^{\circ}$), so most of the
events will survive the cuts and be detected.  Assuming one year
running at luminosity ${\cal L} \sim 20
\text{ fb}^{-1}/\text{yr}$, we expect $\sim 40,000$ events, yielding a
statistical uncertainty of $\sim 0.5\%$. This is further reduced for
longer runs, or if an $\emem$ luminosity comparable to the design
$\epem$ luminosity may be achieved.

2. Backgrounds: Background from electron pair production may be
effectively removed by an acoplanarity cut.  The major remaining
background is then $e^- \nu W^-$ when followed by $W^- \to e^-\nu_e$,
which results from $e_L^-$ contamination in the $e^-_R$ polarized
beams. The cross section for this background is 400 (43) fb for LL
(LR) beam polarization~\cite{Cuypers}. If both beams are 90\%
right-polarized, {\em i.e.}, if only 10\% of the electrons in each
beam are left-handed, the background is reduced to 12 fb. In principle
these backgrounds are calculable and can be subtracted, so the induced
uncertainty in $\sigma_R$ should be negligible.

3. Experimental systematic errors: These include uncertainties in
various collider parameters, including the beam energy, luminosity,
and so on. Accurate knowledge of the beam polarization is also
required.  Note, however, that if beam polarization is a dominant
source of uncertainty, one may use unpolarized beams instead and run
below the $\selectron_L$ pair production threshold for a longer time
to compensate the loss in cross section. The resulting increase in
background is acceptable if well-understood. To compare the
theoretical cross section and the total number of events, detailed
Monte Carlo simulations incorporating effects ranging from initial
state radiation and beamstrahlung to detector acceptances must be
performed to obtain the predicted number of events passing the
cuts. Such simulations are beyond the scope of this paper.  We will
see, however, that experimental systematic errors are likely to be
some of the dominant errors in this analysis, and further studies are
necessary.

After obtaining the cross section $\sigma_R$ from experiment, we need
to extract $h_1$ from $\sigma_R$.  The associated uncertainties
include:

1. Radiative corrections: At the level of precision we are
considering, radiative corrections to the cross section must be
included. These are required to set the low scale $m$ so that the
heavy scale $M$ may be inferred from the measured value of $\susyU_i$.
However, these corrections are calculable, and we expect the
uncertainty to be small after the one-loop radiative corrections are
included.  We have not included such corrections in our calculations.

2. Lepton flavor violation: Until now we have assumed that lepton
flavor is conserved, as is approximately true in a wide variety of
models. However, if the slepton mass matrices are not diagonalized in
the same basis as the lepton mass matrix, the lepton flavor mixing
matrix elements will appear at the gaugino vertices. Such mixing may
reduce the $\emem$ selectron pair signal and cause some uncertainties
in determining $h_1$. However, these lepton flavor violating effects
will be well-probed at the same time. For instance, Ref.~\cite{ACFH}
shows that a mixing angle between the first and second generations of
order $\sin \theta_{12}\sim 0.02$ will be probed at the 5$\sigma$
level. The fractional deviation in the $\emem$ cross section is at
most $2\sin^2\theta_{12} \cos^2\theta_{12}$, and so the induced
uncertainty in deviation in $\susyU_1$ is $\le \frac{1}{2} \sin^2
\theta_{12} \cos^2\theta_{12} \sim 2\times 10^{-4}$. If no lepton 
flavor violation is found, the mixing angles are therefore too small
to induce significant uncertainties in $\susyU_1$. On the other hand,
if lepton flavor violating events are discovered, the total three
generation slepton production cross section may be used instead. The
backgrounds will then include all 3 generations of leptons from $W^-$
decay and will be somewhat larger, but from the discussion above, we
know that they are small enough at an $\emem$ collider and can be
calculated anyway. Lepton flavor violation therefore should not pose a
severe problem, and for simplicity in the remaining discussion, we
will assume it is absent.

3. Uncertainties in experimental determination of $m_{\selectron_R}$
and $m_{\LSP}$: These two masses are the major parameters on which
$\sigma_R$ depends in the gaugino region, and therefore must be known
well for a precise prediction of $\sigma_R$ to be possible. For
simplicity, we assume here that $\LSP$ is pure Bino and
$m_{\LSP}=M_1$; the complication of neutralino mixings will be
discussed next.  The masses $m_{\selectron_R}$ and $m_{\LSP}$ can be
determined from the energy spectrum of the final state electrons in
the $\selectron_R \to e\LSP$ decay. The energy distribution is flat
for two-body decay with two sharp endpoints determined by
$m_{\selectron_R}$, $m_{\LSP}$, and $s$:
\begin{equation}
E_{\rm min}=\frac{m_{\selectron_R}}{2} \left(1-
  \frac{m_{\LSP}^2}{m_{\selectron_R}^2} \right)
  \gamma (1-\beta)\ , \quad
E_{\rm max}=\frac{m_{\selectron_R}}{2} \left(1-
  \frac{m_{\LSP}^2}{m_{\selectron_R}^2} \right)
  \gamma (1+\beta)\ ,
\end{equation}
where 
\begin{equation}
\gamma=\frac{\sqrt{s}}{2m_{\selectron_R}}\ , \quad
\beta=\sqrt{1-\frac{4m_{\selectron_R}^2}{s}} \ .
\end{equation}
One can therefore extract $m_{\selectron_R}$ and $m_{\LSP}$ from
measurements of $E_{\rm min}$ and $E_{\rm max}$. As we will see, the
uncertainties in $m_{\selectron_R}$ and $m_{\LSP}$ are
positively-correlated and form a narrow ellipse-like region in the
$(m_{\selectron_R}, m_{\LSP})$ plane.  At the same time, the
$t$-channel mass insertion implies that, while the total cross section
$\sigma_R$ increases as $m_{\selectron_R}$ decreases, it also
increases as $m_{\LSP}$ increases, and so the constant $\sigma_R$
contours are approximately parallel to the major axis of the ellipse.

\begin{figure}
\centerline{\psfig{file=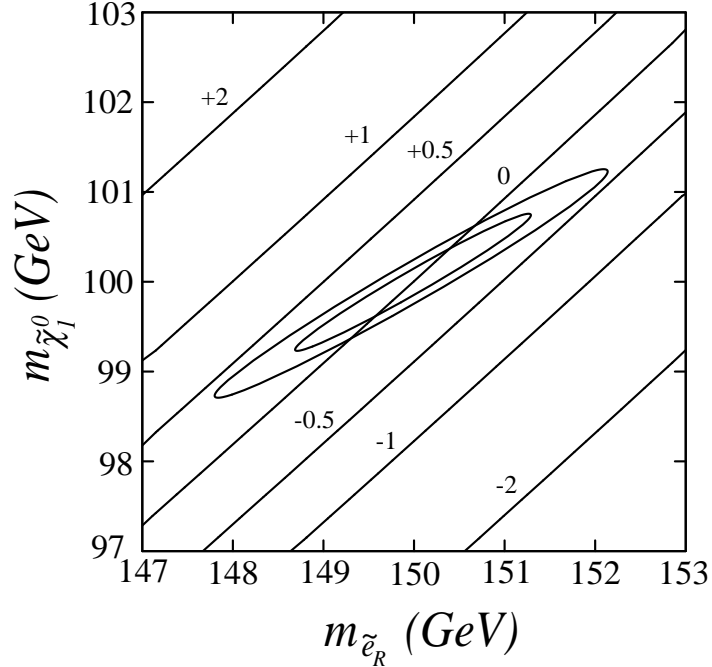,width=0.6\textwidth}}
\vspace*{.15in}
\caption{The allowed regions, ``uncertainty ellipses,'' of the 
($m_{\selectron_R}$, $m_{\LSP}$) plane, determined by measurements of
the end points of final state electron energy distributions with
uncertainties $\Delta E= 0.3$ GeV and 0.5 GeV.  The underlying central
values are $(m_{\selectron_R}, m_{\LSP}) = (150 \text{ GeV}, 100
\text{ GeV})$, and $\protect\sqrt{s}=500$ GeV.  We also superimpose 
contours (in percent) of the fractional variation of $\sigma_R$ with
respect to its value at the underlying parameters. 
\label{fig:ellipse}}
\end{figure}

\begin{figure}
\centerline{\psfig{file=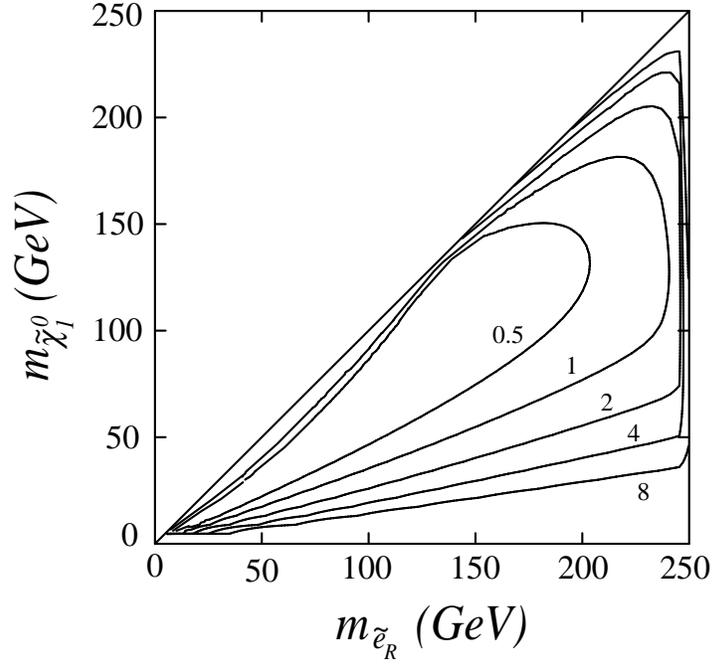,width=0.6\textwidth}}
\vspace*{.15in}
\caption{Contours in the ($m_{\selectron_R}$, $m_{\LSP}$) plane of 
maximal fractional variation in $\sigma_R$ (in percent) on the $\Delta
E=0.3$ GeV ``uncertainty ellipse,'' with $\protect\sqrt{s} =500$ GeV.
\label{fig:variation500}}
\end{figure}

\begin{figure}
\centerline{\psfig{file=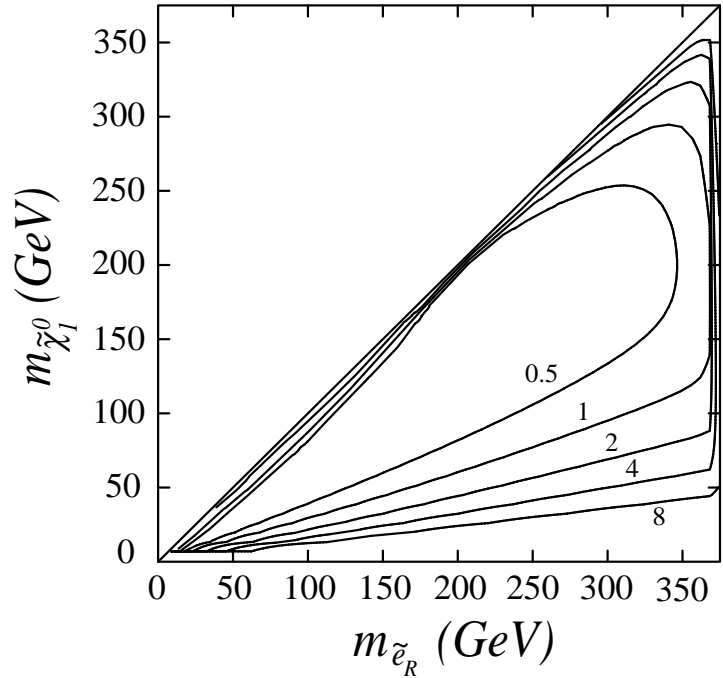,width=0.6\textwidth}}
\vspace*{.15in}
\caption{The same as in Fig.~\protect\ref{fig:variation500}, but
with $\protect\sqrt{s}=750$ GeV.  Note the different scales of the
axes relative to Fig.~\protect\ref{fig:variation500}.
\label{fig:variation750}}
\end{figure}

The variation in $\sigma_R$ on the ``uncertainty ellipse'' can be very
small for some values of $m_{\selectron_R}$ and $m_{\LSP}$. To show
this, we assume that $E_{\rm min}$ and $E_{\rm max}$ are determined
independently with uncertainty $\Delta E$.  The allowed region in the
$(E_{\rm min}, E_{\rm max})$ plane is therefore an ``uncertainty
circle'' with radius $\Delta E$.  This ``uncertainty circle''
transforms into an ``uncertainty ellipse'' in the $(m_{\selectron_R},
m_{\LSP})$ plane, which is shown in Fig.~\ref{fig:ellipse} for the
central values $m_{\selectron_R}=150$ GeV and $m_{\LSP}=100$ GeV. The
$\Delta E= 0.5$ GeV and 0.3 GeV ellipses roughly correspond to the
$\Delta \chi^2 = 4.61$ (90\% C.L.) and $\Delta \chi^2=2.28$ (68\%
C.L.) ellipses given in Ref.~\cite{JLC} for a similar analysis with
smuon pairs.\footnote{We expect our estimates of endpoint energy
uncertainties to be conservative, as they are based on $\epem$ mode
event rates, whereas, given the $\emem$ cross section, data from the
$\emem$ mode should reduce these errors significantly.  The
uncertainties are in fact controlled by a number of factors, including
total cross section, detector energy resolution, electron energy bin
size, and of course, the underlying selectron and neutralino masses.
See Ref.~\cite{slepton} for a discussion of this issue.}  We also
superimpose constant $\sigma_R$ contours on the same figure. We see
that the variation in $\sigma_R$ induced by uncertainties in
$m_{\selectron_R}$ and $m_{\LSP}$ is less than 0.3\% for $\Delta E =
0.3$ GeV and this set of the parameters.  In
Fig.~\ref{fig:variation500}, we show the maximal variations in
$\sigma_R$ in the corresponding $\Delta E = 0.3$ GeV ellipses for
different central values of $m_{\selectron_R}$, $m_{\LSP}$. For
$m_{\LSP}$ not too small and $m_{\selectron_R}$ not too close to
threshold, there is a large region in the $(m_{\selectron_R},
m_{\LSP})$ parameter space in which the variation is less than 1\%,
the target precision. If the variation is too large because
$m_{\selectron_R}$ is too close to threshold, the result can be
improved by raising the beam energy, as shown in
Fig.~\ref{fig:variation750}. The reduction of these theoretical
systematic uncertainties is a great advantage for the precision
measurement of $h_1$ at $\emem$ colliders. In contrast, for $\epem
\to \selectron_R^+ \selectron_R^-$, the constant cross section
contours run roughly perpendicular to the uncertainty ellipse,
resulting in a much larger uncertainty.

\begin{figure}
\centerline{\psfig{file=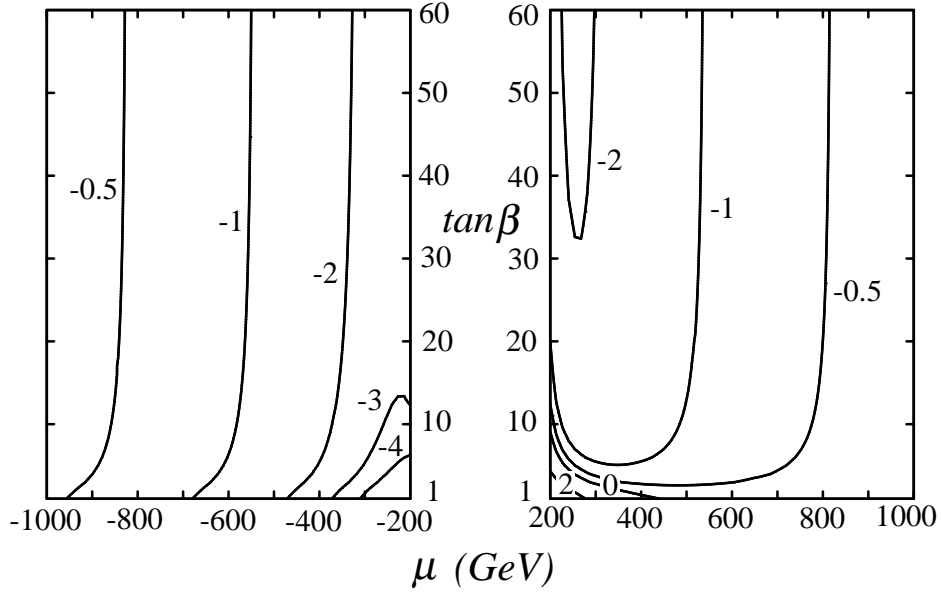,width=0.8\textwidth}}
\vspace*{.15in}
\caption{The fractional variation in $\sigma_R$ (in percent) in the 
($\mu$, $\tan\beta$) plane, with respect to the $\mu\to \infty$ limit,
for $(m_{\selectron_R}, m_{\LSP})= (150 \text{ GeV}, 100\text{ GeV})$,
with $\protect\sqrt{s}=500$ GeV.  $M_2$ is assumed to be $2M_1$, and
for each point in the plane, their values are fixed by $m_{\LSP}$.
\label{fig:mutanbeta}}
\end{figure}

4. Neutralino mixings: In the discussion so far, we have assumed that
the lightest neutralino is pure Bino. This is only true in the limit
of $|\mu| \to \infty$. A general neutralino mass matrix depends on the
four parameters $M_1$, $M_2$, $\mu$, and $\tan \beta$. To correctly
calculate the cross section, one has to diagonalize the neutralino
mass matrix and include contributions from all four neutralino mass
eigenstate propagators.  Although the dependence of $\sigma_R$ on
$M_2$, $\mu$, and $\tan\beta$ should be weak in the gaugino region,
they are not negligible at the required level of precision. To
investigate this, we have calculated $\sigma_R$ for different choices
of $M_1$, $M_2$, $\mu$, and $\tan\beta$ while keeping the measurable
$m_{\LSP}$ fixed. By explicit calculation we find that the dependence
on $M_2$ of $\sigma_R$ is very weak, since $\tilde{B}$ and
$\tilde{W_3}$ only mix indirectly, and the variation in $\sigma_R$ is
much smaller than 1\% for reasonable variations in $M_2$. We may
therefore assume $M_2 = 2 M_1$ without loss of generality.  In
Fig.~\ref{fig:mutanbeta} we show the fractional variation of
$\sigma_R$ relative to the pure Bino limit as a function of $\mu$ and
$\tan\beta$ for fixed $m_{\LSP}=100 \text{ GeV}$ and
$m_{\selectron_R}=150 \text{ GeV}$. The value of $M_2=2M_1$ is
determined by requiring the correct value of $m_{\LSP}$.  We see that
the variation of $\sigma_R$ is small for large $|\mu|$ (less than 1\%
for $\mu \agt 500\text{ GeV}$ or $\mu \alt -600 \text{ GeV}$) but can
be up to 2--4\% for smaller $|\mu|$.  Therefore, in order to be able
to calculate $\sigma_R$ at the 1\% level, some information about $\mu$
and $\tan\beta$ is needed: either a lower bound of $|\mu|\agt 500-600
\text{ GeV}$ is required, or $\mu$ and $\tan\beta$ must be bounded to
lie within a certain range if the underlying value of $|\mu|$ is
smaller. Such bounds may be obtained from some other processes in
different colliders.  For example, $\tilde{\chi}_1^0 \tilde{\chi}_3^0$
production (in $\epem$ collisions) may probe $\mu$ up to
$\sqrt{s}-m_{\LSP}$. Energies of $\sqrt{s}\sim 1
\text{ TeV}$, if available, will therefore allow either a
determination of $\mu$ or a sufficiently high lower bound on $\mu$ for
us to obtain a precise prediction of $\sigma_R$ so that $h_1$ can be
extracted with small uncertainties.

Finally, many of the above considerations apply also to left-handed
selectrons.  If kinematically accessible, their production cross
section $\sigma_L$ at $\emem$ colliders may also be used to precisely
measure gaugino couplings, since the $\selectron_L^-
\selectron_L^-$ pair production cross section receives contributions
from both $t$-channel $\tilde{B}$ and $\tilde{W}^3$ exchange, and
hence depends on both $h_1$ and $h_2$. For equivalent mass selectrons,
$\sigma_L$ is generally even larger than $\sigma_R$.  Note also that
$\selectron_L$ and $\selectron_R$ production may be separated either
by beam polarization, or, if the selectrons are sufficiently
non-degenerate, by kinematics~\cite{JLC} or by running below the
higher production thresholds.  If the $\charginoone$ and
$\neutralinotwo$ decay channels are not open, the only decay is
$\selectron_L^-\to e^-\LSP$ and we will have a large clean sample of
events for precision studies.  However, in general, the decay patterns
may complicate the analysis.  The cross section also depends strongly
on $m_{\neutralinotwo}$ (in the gaugino region), which could be
measured either directly from $\neutralinotwo \neutralinotwo$
production in $\epem$ collisions, or indirectly by measuring $M_1$,
$M_2$, $\mu$ and $\tan\beta$ from chargino and $\LSP$ properties. In
the end, a measurement of $\sigma_L$ bounds a certain combination of
$h_1$ and $h_2$. Under the assumption that the heavy sparticles are
fairly degenerate, the deviations $\susyU_1$ and $\susyU_2$ are
related and determined by the same heavy scale $M$, and so $\sigma_L$
also provides a probe of the heavy scale $M$, which, in fact, is
generically more sensitive, since $\susyU_2 > \susyU_1$ in most
models.  Of course, in the event that both $\selectron_R$ and
$\selectron_L$ are studied, both $\susyU_1$ and $\susyU_2$ may be
determined, and we may check that their implications for the heavy
scale $M$ are consistent or find evidence for non-degeneracies in the
heavy sector.

In summary, we find that for a fairly general region of the parameter
space, selectron production at an $\emem$ collider may provide an
extremely high precision measurement of the gaugino coupling $h_1$ and
super-oblique parameter $\susyU_1$.  We have investigated both
experimental statistical and theoretical systematic uncertainties.  By
exploiting many appealing features of the $\emem$ mode, most
uncertainties may be reduced to below 1\% in the cross section
measurement.  The dominant theoretical systematic uncertainty appears
to be from neutralino mixings, but even these may be reduced below the
1\% level with information from other processes.  The remaining
uncertainties are experimental systematic uncertainties.  These
include, for example, the luminosity uncertainty, which has been
estimated to be $\sim 1\%$~\cite{JLC}.  Such issues require further
study. Nevertheless, the $\emem$ mode certainly appears more promising
than the $\epem$ mode.  If such errors may be reduced to the 1\%
level, a precision measurement of $\susyU_1$ at the level of $0.3\%$
will be possible, providing not only a stringent test of SUSY, but
also allowing us to bound the mass scale of the heavy sector to within
a factor of 3, even if they are beyond the reach of the LHC.  Such a
stringent bound would provide strong constraints for model-building,
and, in the most optimal case, would provide a target for sparticle
searches at even higher energy colliders.

\section{Probe of SU(3) Couplings from squarks}
\label{sec:squarks}

In this section, we consider the possibility of probing the heavy
superparticle mass scale through their effects on SU(3) gluon and
gluino couplings.  Such probes require that strongly-interacting
sparticles be accessible.  Such is the case in the 2--1 models
discussed in Sec.~\ref{sec:introduction}, and these are the scenarios
we will consider here.  The most relevant decoupling parameters for
our study below will be $\dsusyU_{32}$ and $\dsusyU_{31}$.  In 2--1
models,

\begin{equation}
\dsusyU_{32(31)} = \frac{h_3/h_{2(1)}}{g_3/g_{2(1)}}-1 \approx
1.8\% (2.2\%) \times \ln \frac{M}{m} \ .
\end{equation}
For heavy superpartners at $M \sim {\cal O}(10 \text{ TeV})$, these
corrections can be as large as $\sim 10\%$, much larger than for the
corresponding SU(2) and U(1) couplings, and so are promising to
investigate.

In 2--1 models, the gluino and third generation sfermions are light,
but all other sfermions are heavy.  SU(3) effects may then be measured
in processes involving gluinos and the bottom and top squarks.  At
$\epem$ colliders, squarks may be pair-produced in large
numbers~\cite{Don,Bartl}.  However, squark pair production takes place
only through $s$-channel $\gamma$ and $Z$ processes, and so is
independent of $h_i$.  To find cross sections that do depend on $h_3$,
one may turn to three-body processes, such as $b\tilde{b}\tilde{g}$
and $t\tilde{t}\tilde{g}$, as was noted in Sec.~\ref{sec:observables}.
In this section, however, we will focus on another possibility and
consider measurements of $h_3$ through squark decay branching ratios.

Any of the $\tilde{b}_{L,R}$ and $\tilde{t}_{L,R}$ squarks may be used
as a probe.  However, the decay paths and backgrounds vary greatly
depending on the particular mass patterns of these squarks and the
gluino.  The boundary conditions for the light sparticle masses are
not in general universal, and this is in fact the underlying
motivation for the 2--1 framework.  The low-energy spectrum may
therefore be arbitrary, although, of course, the $\tilde{t}_{L}$ and
$\tilde{b}_L$ masses are still related by SU(2) invariance.  For
concreteness, we will primarily focus on $\tilde{b}_L$ decays.  As
will be described below, our analysis will rely only on the number of
events with 3 or more tagged $b$ jets.  For simplicity, we will assume
that the contributions of other third generation squarks to such
events are negligible.  This is the case either if these squarks are
too heavy to be produced, or if their masses are such that their
decays to gluinos are closed or highly phase-space suppressed. (Note
that top squark decays to gluinos are also suppressed by the large top
quark mass.)  We also take the left-right mixing in the $\tilde{b}$
sector to be negligible.  Such an assumption may be tested by
measurements of the $\tilde{b}_L$ properties themselves~\cite{Bartl},
or, for example, by measurements of $\tan\beta$ from other
sectors~\cite{Moroi}.  Finally, we assume that the lighter neutralinos
and chargino are well-studied and are determined to be highly
gaugino-like by, for example, directly measuring or 
placing lower bounds on Higgsino
masses.\footnote{If, however, the Higgsinos are in the heavy sector,
significant non-decoupling contributions to the gaugino couplings from
the large third generation Yukawa couplings must be
included~\cite{CFP}.}

As individual decay widths are difficult to measure, our analysis will
depend on measuring branching ratios, and is only possible when two or
more decay modes are open.  As we are interested in the SU(3) gaugino
coupling $h_3$ in this section, we assume $m_{\tilde{g}} + m_b <
m_{\tilde{b}_L}$ so that the gluino decay mode is open. (Of course, if
the gluino decay mode is closed but both Wino and Bino decay modes are
open, a measurement of $h_2/h_1$ from these branching ratios may also
be used to probe decoupling effects.) The branching ratios then depend
on $h_3/h_2$ and $h_3/h_1$ and probe the decoupling parameters
$\dsusyU_{32 (31)}$ given above.

The two-body decay widths of $\tilde{b}_L$ to $\tilde{g}$, 
$\charginoone$, $\neutralinotwo$, and $\LSP$ (assuming
that $\charginoone$, $\neutralinotwo$, and $\LSP$ are
pure gauginos) are
\begin{eqnarray}
\Gamma (\tilde{b}_L \to b \tilde{g}) &=& \frac{4}{3}
\frac{h_3^2}{8\pi} m_{\tilde{b}_L} P( m_{\tilde{b}_L},
m_{\tilde{g}}, m_b) \equiv 
\frac{h_3^2}{8\pi} m_{\tilde{b}_L} P_3\ , \nonumber \\
\Gamma (\tilde{b}_L \to b \neutralinotwo) &=& \frac{1}{4}
\frac{h_2^2}{8\pi} m_{\tilde{b}_L} P( m_{\tilde{b}_L},
m_{\neutralinotwo}, m_b) \equiv 
\frac{h_2^2}{8\pi} m_{\tilde{b}_L} P_2\ , \nonumber \\
\Gamma (\tilde{b}_L \to t \charginoone) &=& \frac{1}{2}
\frac{h_2^2}{8\pi} m_{\tilde{b}_L} P( m_{\tilde{b}_L},
m_{\charginoone}, m_t) \equiv 
\frac{h_2^2}{8\pi} m_{\tilde{b}_L} {P_2}'\ , \nonumber \\
\Gamma (\tilde{b}_L \to b \LSP) &=& \frac{1}{36}
\frac{h_1^2}{8\pi} m_{\tilde{b}_L} P( m_{\tilde{b}_L},
m_{\LSP}, m_b) \equiv 
\frac{h_1^2}{8\pi} m_{\tilde{b}_L} P_1\ ,
\end{eqnarray}
where these equations define $P_3$, $P_2$, $P'_2$, and $P_1$, and

\begin{eqnarray}
P(m_0, m_1, m_2) &=& \theta (m_0-m_1-m_2) \left\{  \left( 1-
\frac{m_1^2+m_2^2}{m_0^2}\right) \sqrt{\left( \frac{1}{2} - 
\frac{m_1^2}{2m_0^2} +\frac{m_2^2}{2m_0^2}\right)^2 - 
\frac{m_2^2}{m_0^2}} \right.\nonumber \\
&&\left. + 2 \left[ \left(\frac{1}{2}-\frac{m_1^2}{2m_0^2}
+\frac{m_2^2}{2m_0^2}\right)^2 - 
\frac{m_2^2}{m_0^2}\right]  \right\} 
\end{eqnarray}
is the phase space factor for a scalar particle of mass $m_0$ decaying
into two fermions with masses $m_1$ and $m_2$.  The branching ratio
for $\tilde{b}_L \to b\gluino$ is then given by

\begin{equation}
B_{\gluino}= B(\tilde{b}_L \to b\gluino) =
\frac{D_{32}^2 P_3}{D_{12}^2 P_1 +P_2 +{P_2}' +D_{32}^2 P_3} \ ,
\label{eq:branchingratio}
\end{equation}
where $D_{ij}\equiv h_i/h_j =(1+\dsusyU_{ij}) g_i/g_j$.

The deviation of $D_{12}=h_1/h_2$ from $g_1/g_2$ is much smaller than
that of $D_{32}$ from $g_3/g_2$, and the term involving $D_{12}$ is
suppressed by the small U(1) coupling, so it is a good approximation
to fix $D_{12}=g_1/g_2$. If the gluino branching fraction can be
measured, and all the relevant particle masses are known, then from
Eq.~(\ref{eq:branchingratio}) we can obtain $D_{32}$:

\begin{equation}
D_{32}= \left[ \frac{D_{12}^2 P_1 +P_2 +{P_2}'}{P_3} 
\frac{B_{\gluino}}{1-B_{\gluino}} \right]^{\frac{1}{2}} \ .
\label{eq:masterformula}
\end{equation}
Combining this with the measured value of $g_3/g_2$\footnote{Assuming
that the ${\cal O}(\alpha_s^3)$ perturbative QCD corrections are
calculated, the uncertainty in $\alpha_s(m_Z^2)$ from $q\bar{q}$
events at the NLC is estimated to be at the 1\% level~\cite{NLC} and
is therefore negligible for this study.}, we then have a measurement
of $\dsusyU_{32}$ and a constraint on the heavy sector mass scale.  Of
course, as in the previous sections, such a measurement is subject to
a number of uncertainties. Uncertainties in the measurement of
$B_{\gluino}$ arise from statistical fluctuations, backgrounds, and
experimental systematic errors, while the extraction of $D_{32}$ from
$B_{\gluino}$ is subject to theoretical systematic uncertainties from
imprecisely known SUSY parameters. We will discuss the theoretical
systematic uncertainties first.

The major theoretical systematic uncertainties are the uncertainties
in $m_{\tilde{b}_L}$ and $m_{\gluino}$.  For all measurement methods,
these masses enter the determination of $D_{32}$ through the phase
space factors in Eq.~(\ref{eq:masterformula}). In addition, depending
on the method used to measure $B_{\gluino}$, a dependence on
$m_{\tilde{b}_L}$ may also enter through this quantity.  This is the
case, for example, if $B_{\gluino}$ is determined by comparing the
number of events in a particular channel to the total
$\tilde{b}_L\tilde{b}_L$ cross section, and this total cross section
is determined theoretically by its dependence on $m_{\tilde{b}_L}$.
However, the uncertainties entering from the dependence of
$B_{\gluino}$ on $m_{\tilde{b}_L}$, in addition to being
method-dependent, are typically negligible relative to other errors.
For example, for the method just described, we have found that for
$m_{\tilde{b}_L}$ significantly below threshold, the uncertainty from
the dependence of $B_{\gluino}$ on $m_{\tilde{b}_L}$ is small compared
to that from the phase space factors.  This is no longer the case for
$m_{\tilde{b}_L}$ near threshold, as there the total cross section is
sensitive to $m_{\tilde{b}_L}$, but in this region, the cross section
is small and statistical uncertainties are dominant.

\begin{figure}
\centerline{\psfig{file=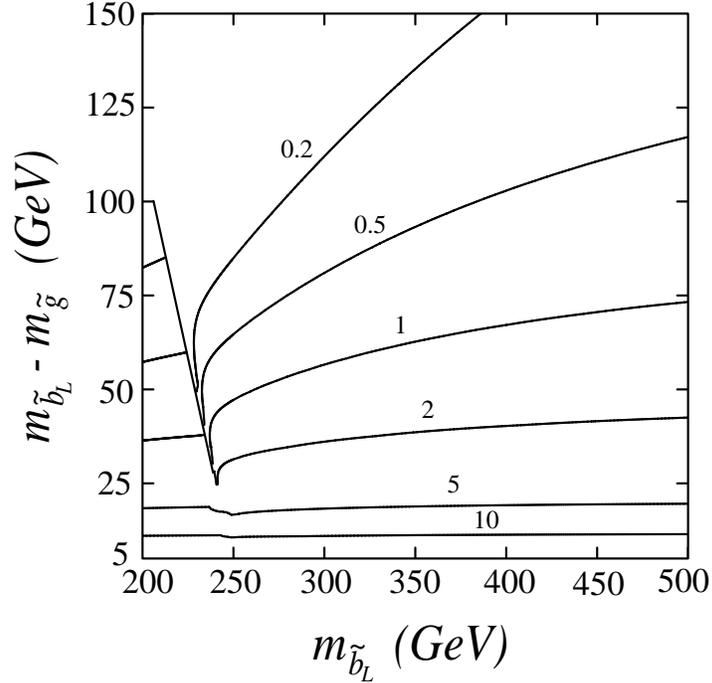,width=0.6\textwidth}}
\vspace*{.15in}
\caption{Systematic uncertainty in $\dsusyU_{32}$ 
arising from uncertainty in $m_{\tilde{b}_L}$.  Plotted are contours
of constant variation in $\dsusyU_{32}$ per GeV variation in
$m_{\tilde{b}_L}$, $\frac{\Delta \dsusyU_{32}}{\Delta
m_{\tilde{b}_L}}$, in percent for $\protect\sqrt{s}=1$ TeV in the
$(m_{\tilde{b}_L}, m_{\tilde{b}_L}-m_{\tilde{g}})$ plane. Some
regions of this plane correspond to gluino masses that are already
excluded by current bounds.
\label{fig:msbottom}}
\end{figure}

\begin{figure}
\centerline{\psfig{file=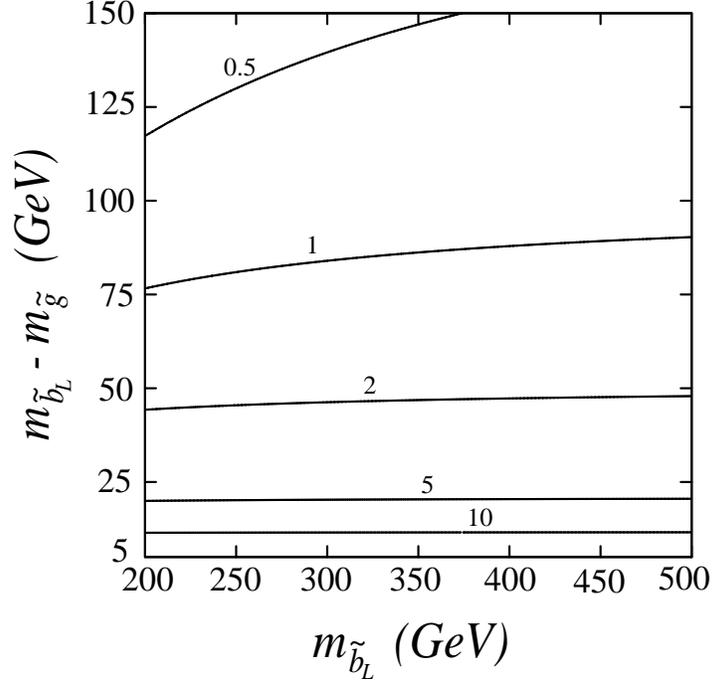,width=0.6\textwidth}}
\vspace*{.15in}
\caption{Systematic uncertainty in $\dsusyU_{32}$ arising from 
uncertainty in $m_{\tilde{g}}$.  Plotted are contours of constant
variation in $\dsusyU_{32}$ per GeV variation in $m_{\tilde{g}}$,
$\frac{\Delta \dsusyU_{32}}{\Delta m_{\tilde{g}}}$, in percent in the
$(m_{\tilde{b}_L}, m_{\tilde{b}_L}-m_{\tilde{g}})$ plane for
$\protect\sqrt{s}=1$ TeV.
\label{fig:mgluino}}
\end{figure}

We therefore consider only the theoretical systematic uncertainties
from the phase space factors. The fractional uncertainties in
$D_{32}$, or equivalently, the uncertainties in $\dsusyU_{32}$, from
$m_{\tilde{b}_L}$ and $m_{\gluino}$ systematic errors are given by

\begin{eqnarray}
\frac{d \dsusyU_{32}}{d m_{\tilde{b}_L}} &\approx &
\frac{1}{D_{32}} \frac{d D_{32}}{d m_{\tilde{b}_L}} =
\frac{1}{2(D_{12}^2 P_1 +P_2 +{P_2}')} 
\left(D_{12}^2 \frac{\partial P_1}{\partial
m_{\tilde{b}_L}} +\frac{\partial P_2}{\partial
m_{\tilde{b}_L}} +\frac{\partial {P_2}'}{\partial
m_{\tilde{b}_L}}\right) -
\frac{1}{2 P_3} \frac{\partial P_3}{\partial
m_{\tilde{b}_L}} \\
\frac{d \dsusyU_{32}}{d m_{\gluino}} &\approx &
\frac{1}{D_{32}} \frac{d D_{32}}{d m_{\gluino}} =
-\frac{1}{2 P_3} 
\frac{\partial P_3}{\partial
m_{\gluino}} \ .
\end{eqnarray}
We plot the systematic errors from $m_{\tilde{b}_L}$ and $m_{\gluino}$
in Figs.~\ref{fig:msbottom} and \ref{fig:mgluino}, respectively. The
uncertainties in $\dsusyU_{32}$ are in percent per GeV variation in
$m_{\tilde{b}_L}$ or $m_{\gluino}$ and are plotted in the
$(m_{\tilde{b}_L}, m_{\tilde{b}_L} - m_{\gluino})$ plane. Motivated by
the current bounds on squark masses and the prejudice that colored
superparticles should be heavier than uncolored ones, we have taken a
value of $\sqrt{s} = 1 \text{ TeV}$ such that we may pair produce
squarks with masses of up to 500 GeV.  (Note that some regions of the
plane are for gluino masses that have already been excluded by current
bounds.)  At each point, we have assumed that the underlying
parameters are given by the gaugino mass unification relations
$m_{\gluino} = 3.3 m_{\neutralinotwo} = 3.3 m_{\charginoone} = 6.6
m_{\LSP}$.  (The abrupt behavior of the contours in
Fig.~\ref{fig:msbottom} results from the opening of the decay
$\tilde{b}_L \to t \charginoone$.)  For decreasing
$m_{\tilde{b}_L}-m_{\gluino}$, the uncertainties increase, as the
phase space for decays to gluinos shrinks and the decay width to
gluinos becomes more sensitive to $m_{\tilde{b}_L}$ and $m_{\gluino}$.
The theoretical systematic uncertainty in $\dsusyU_{32}$ is therefore
highly dependent on the mass splitting $m_{\tilde{b}_L}-m_{\gluino}$.
We see generally, however, that for this uncertainty to be below 10\%,
$m_{\tilde{b}_L}$ and $m_{\gluino}$ typically must be measured to
within a few GeV. Measurements of squark masses at this level have
been shown to be possible at the NLC, even in the presence of cascade
decays~\cite{Don}.  Gluino masses may be measured at the NLC in the
scenarios we are considering through squark decays to gluinos.
Alternatively, it is possible that the mass difference
$m_{\tilde{b}_L}-m_{\gluino}$ could be measured at the LHC through
methods similar to those described in Ref.~\cite{SnowmassLHC}.
However, estimates of the gluino mass resolution certainly merit
further investigation.

The phase space factors also depend on other mass parameters as well,
such as $m_{\tilde{\chi}^0_{1, 2}}$ and $m_{\charginoone}$, so there
are also uncertainties induced by these unknown masses. However, these
masses are expected to be much smaller than $m_{\tilde{b}_L}$ and
$m_{\gluino}$. The phase space factors are therefore larger for
$\tilde{b}_L$ decays into these particles and are less sensitive to
their masses. In addition, these masses will probably be known more
precisely than $m_{\tilde{b}_L}$ and $m_{\gluino}$. We therefore
expect the uncertainties coming from these other masses to be much
smaller than those from $m_{\tilde{b}_L}$ and $m_{\gluino}$.

In the above discussion, we assume that the lighter neutralinos and
charginos are pure gauginos. As discussed in the previous sections,
neutralino and chargino mixings may also introduce some uncertainties
in determining the gaugino couplings. However, here the non-decoupling
effects we expect are much larger ($\sim 10\%$ versus $\sim 1-3\%$ in
previous cases). The uncertainties from these mixings, while possibly
significant for the previous cases, are expected to be small relative
to the 10\% corrections possible in the SU(3) couplings.

We now consider the experimental statistical and systematic errors
arising in the measurement of $B_{\gluino}$.  To measure this
branching fraction, we will exploit the fact that gluino decays tend
to give more $b$ quarks in the final state than do decays to the
electroweak gauginos. Decays to the Bino and Winos produce one $b$
quark.  Decays to gluinos are followed by gluino decays, which in 2--1
models are dominated by decays through off-shell $t$- and $b$-squarks,
resulting in an additional two $b$ quarks in the final
state.\footnote{In fact, additional $b$ quarks may appear in both Wino
and gluino decay modes if neutralinos $\neutralinotwo$ are produced
that then decay via $\neutralinotwo \to b\bar{b}\LSP$.  We will assume
that this $\neutralinotwo$ branching fraction is well-measured.  For
simplicity, in the quantitative results presented below, we assume
that $\neutralinotwo$ decays to $b$ quarks are absent, as would be the
case, for example, if the two-body decay $\neutralinotwo \to
\tilde{\tau} \tau$ is open.}  Thus, $\tilde{b}_L \tilde{b}_L$ pair
events with 0, 1, and 2 gluino decays result in 2, 4, and 6 $b$
quarks, respectively.  

At the NLC, excellent $b$-tagging efficiencies and purities are
expected.  We will take the probability of tagging a $b$ ($c$) quark
as a $b$ quark to be $\epsilon_b = 60\%$ ($\epsilon_c = 2.6\%$), with
a negligible probability for light quarks~\cite{Jackson}.  We also
make the crude assumption that the probability of tagging multi-$b$
events is given simply by combinatorics, so that the probability of
tagging $m$ of $n$ $b$ jets is ${n\choose m} \epsilon_b^m
(1-\epsilon_b)^{n-m}$.  With these assumptions, we may bound the
gluino branching fraction by measuring $N_i$, the number of events
with $i = 3,4,5$ {\em tagged} $b$ jets, along with the total cross
section determined by $m_{\tilde{b}_L}$, which we assume is measured
by kinematical arguments~\cite{Don}. (We may also use other channels;
however, $N_2$ receives huge backgrounds from $t\bar{t}$ production,
and the number of events with 6 tagged $b$ jets is not statistically
significant.)  The standard model backgrounds to multi-$b$ events
include $t\bar{t}$, $t\bar{t}Z$, $ZZZ$, $\nu\bar{\nu}ZZ$, and
$t\bar{t}h$~\cite{HMthesis}.  At 1 TeV, the resulting backgrounds with
3, 4, and 5 tagged $b$ jets, after including all branching fractions
and the tagging efficiencies given above, were calculated in
Ref.~\cite{Moroi} and found to be 4.0 fb, 1.0 fb, and 0.0095 fb,
respectively. In our calculations we include only standard model
backgrounds. Additional multi-$b$ events may arise from other SUSY
processes, such as $\tilde{t}\tilde{t}$ production followed by decays
$\tilde{t} \to t\tilde{g}$. Such squark processes also are dependent
on the super-oblique parameters, however, and so may be included as
signal. The analysis will be more complicated and will not be 
considered here.

\begin{figure}
\centerline{\psfig{file=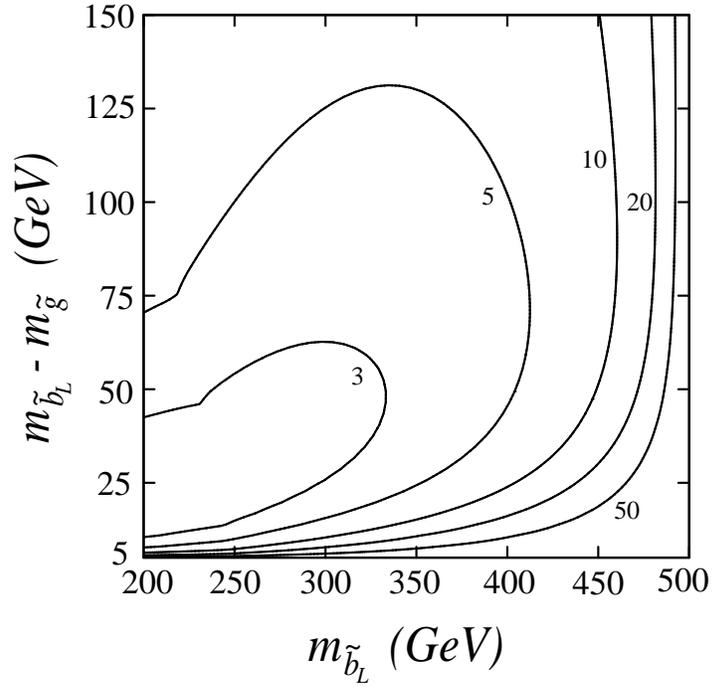,width=0.6\textwidth}}
\vspace*{.15in}
\caption{The error in $\dsusyU_{32}$ in percent
from statistical uncertainties in the multiple $b$-tag events in the
$(m_{\tilde{b}_L}, m_{\tilde{b}_L}-m_{\tilde{g}})$ plane, for
$\protect\sqrt{s}=1$ TeV and integrated luminosity $L=200 \text{
fb}^{-1}$. The assumed $b$-tagging efficiency is $\epsilon_b= 60\%$.
\label{fig:blstat}}
\end{figure}

We would now like to determine quantitatively what bounds on
deviations in $\dsusyU_{32}$ may be set by measurements of $N_i$.  We
will take a central value of $\dsusyU_{32}=0$; we expect the errors to
be uniform for other central values.  We define a simple
$\Delta\chi^2$ variable

\begin{equation}
\Delta\chi^2 \equiv \sum_{i=3}^5 \frac{(N_i - N'_i)^2}{N'_i} \ ,
\end{equation}
where $N_i$ is the sum of the number of signal and background events
with $i$ tagged $b$ jets assuming $\dsusyU_{32} = 0$, and $N'_i$
is the similar quantity for a postulated $\dsusyU'_{32}$.  For given
underlying parameters $\sqrt{s}$, $m_{\tilde{b}_L}$, $m_{\gluino}$ and
integrated luminosity $L$, the values of $\dsusyU'_{32}$ yielding
$\Delta\chi^2=1$ (68\% C.L.) then give the statistical uncertainty.
The fractional error in $\dsusyU_{32}$ from such statistical
uncertainties for $\sqrt{s}=1$ TeV and (unpolarized) integrated
luminosity $L=200 \text{ fb}^{-1}$ is given Fig.~\ref{fig:blstat}.
The statistical uncertainties grow rapidly as $m_{\tilde{b}_L}$
approaches its threshold limit of 500 GeV, as expected.  The
statistical uncertainty, however, also depends on the mass difference
$m_{\tilde{b}_L} - m_{\gluino}$.  For optimal mass splittings, the
gluino decay is fairly phase space-suppressed, yielding roughly an
equal number of gluino and Wino decays. The number of events in the
different channels $N_i$ is then highly sensitive to variations in
$\dsusyU_{32}$. However, for large or very small mass splittings,
either the gluino or the Wino decay dominates, in which case
sensitivity to $\dsusyU_{32}$ is weak.

The total error receives contributions from all three of the sources
shown in Figs.~\ref{fig:msbottom}--\ref{fig:blstat}. We see that if
$\tilde{b}_L$ squarks are produced significantly above threshold, the
$\tilde{b}_L$ and gluino masses are measured to a few GeV, and the
squark-gluino mass splittings are moderate, in the range $25\text{
GeV} \alt m_{\tilde{b}_L} - m_{\gluino} \alt 100\text{ GeV}$, the
combined uncertainty is below the $\sim 10\%$ level.  For nearly ideal
mass splittings, the uncertainties can be much below this level,
possibly yielding a precise measurement of the heavy sector scale.
Note, however, that possibly large experimental systematic errors have
not been included. For this study, a particular source of concern is
the $b$ tagging efficiency for multi-$b$ events, which must be
well-understood for an accurate measurement to be possible.

\begin{figure}
\centerline{\psfig{file=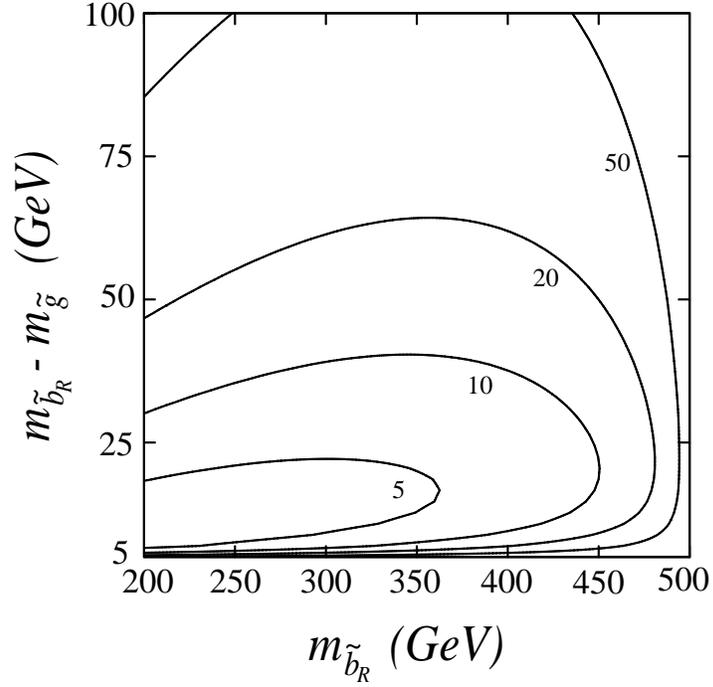,width=0.6\textwidth}}
\vspace*{.15in}
\caption{The error in $\dsusyU_{31}$ in percent from the statistical 
uncertainty in double direct Bino decay events in the
$(m_{\tilde{b}_R}, m_{\tilde{b}_R} - m_{\tilde{g}})$ plane, for
$\protect\sqrt{s}=1$ TeV.  The assumed integrated luminosity is $L=200
\text{ fb}^{-1}$, and the efficiency for the signal is taken to be
$\epsilon= 70\%$; the contours scale as $1/\protect\sqrt{L\epsilon}$.
\label{fig:brstat}}
\end{figure}

Before concluding, we consider briefly the possibility of measuring
the super-oblique parameters through $\tilde{b}_R$ branching ratios.
In this case, the Wino decays are closed, and so only the gluino and
Bino modes compete.  We find that the strongest bounds on
$\dsusyU_{31}$ come from the observation of squark pair events in
which both squarks decay directly to Binos.  Such decays yield clean
events with only two acoplanar $b$ jets, and may be isolated from
standard model backgrounds with simple cuts~\cite{Grivaz}.  In
Ref.~\cite{Don}, such cuts were found to yield efficiencies of
60--80\% for squark pair events.  By measuring the number of double
direct Bino decays, and again determining the total cross section by
measuring $m_{\tilde{b}_R}$ kinematically, bounds on $\dsusyU_{31}$
may be found.  In Fig.~\ref{fig:brstat}, the statistical uncertainties
from such a determination are given.  Not surprisingly, we find that
in this case, a large phase space suppression of the gluino mode is
required to enhance the number of double Bino decay events. A
statistical uncertainty at the level of $\sim 10\%$ is achievable only
for $m_{\tilde{b}_R} - m_{\gluino} \alt 30 \text{ GeV}$. Of course,
one may also include data from multi-$b$ events as in the previous
case, but such considerations do not improve the results noticeably.

\section{Conclusions}
\label{sec:conclusions}

If some of the superpartners of the standard model particles are heavy
and beyond the reach of planned future colliders, we must rely on
indirect methods to study them before their discovery. Such heavy
superpartners decouple from most experimentally accessible
processes. However, heavy superpartner masses break supersymmetry, and
so violate the SUSY relations $g_i=h_i$ between gauge boson and
gaugino couplings at scales below the heavy superpartner mass scale
$M$.  Deviations from these relations are most conveniently
parametrized in terms of the super-oblique parameters
$\widetilde{U}_{i}$ \cite{CFP} and increase logarithmically with $M$.
Therefore, precision measurements of the gaugino couplings $h_i$ in
processes involving the light superpartners will provide important
(and possibly the only) probes of the heavy superpartner sector for
the foreseeable future.

There are many low energy processes and observables involving the
light superpartners and gauginos that depend on the gaugino couplings
$h_i$ and therefore may serve as probes of the super-oblique
parameters.  These were systematically classified in
Sec.~\ref{sec:observables}.  However, in practice, these observables
are subject to many systematic and statistical uncertainties, and not
all of them can be measured to the required precision to provide
significant bounds on the heavy sector. In this paper, we studied
three promising examples at proposed linear $e^{\pm} e^-$ colliders,
one for each of the three coupling constant relations using three
different superparticles processes.  We exploited the versatility of
planned linear colliders, such as their highly polarized beams,
tunable beam energy, and the $e^- e^-$ option, to improve the
precision of the measurements.

In the first example, chargino pair production in $e^+ e^-$ collisions
was used to study the SU(2) gaugino coupling $h_2$. {}From the total
cross section, the truncated forward-backward asymmetry, and a
precisely measured sneutrino mass $m_{\tilde{\nu}_e}$, measurements of
the super-oblique parameter $\susyU_2$ at the level of $\sim 2-3\%$
are possible. We demonstrated the importance of being able to choose
an optimal beam energy so that the experimental observables are most
sensitive to $\susyU_2$.  Note that, since we expect greater
deviations in the SU(2) relation than the U(1) relation, such results
provide bounds on the heavy scale $M$ that are roughly equivalent to
those previously achieved with $\selectron_R$ pair production at
$\epem$ colliders~\cite{slepton}.

In the second example, we considered a measurement of $h_1$ from
$\tilde{e}_R^- \tilde{e}_R^-$ production at an $e^- e^-$ collider.
Such colliders allow measurements that are extremely clean both
experimentally and theoretically, and therefore provide an excellent
environment for precision studies. Such measurements also suffer less
from uncertainties in the relevant SUSY parameters. If the
experimental systematic uncertainties are under control, $\susyU_{1}$
may be measured to $\sim 0.3\%$ for a wide range of the parameter
space. Such a high precision measurement may provide a determination
of the heavy scale within a factor of 3, which is a striking
improvement over the $\epem$ results described above.

The last observables we considered were branching ratios of bottom
squarks decay into gluinos and other gauginos.  These decays can be
used to measure the ratios of gaugino couplings $h_3/h_2$ and
$h_3/h_1$. Although larger uncertainties are usually associated with
strongly-interacting particles, the deviation from the SUSY relation
$h_3=g_3$ is also expected to be larger.  We find that, for squark
production significantly above threshold and small to moderate
squark-gluino mass splittings, it is possible to obtain a measurement
of $\dsusyU_{32}$ which is sensitive to deviations from the SUSY
relation.

These examples imply that the prospects for precision measurements of
gaugino couplings in different scenarios are indeed promising. We have
studied various possible uncertainties in these measurements and find
that most of them may be controlled (at least in some region of the
parameter space), though a complete understanding of all uncertainties
would require detailed experimental simulations that are beyond the
scope of this study.  For this study, it is crucial that collider
parameters be well understood and precisely measured.  Further
experimental studies on these issues are strongly encouraged.

The implications of measurements of the super-oblique parameters
depend strongly on what scenario is realized in nature.  If some
number of superpartners are not yet discovered, bounds on the
super-oblique parameters may lead to bounds on the mass scale of the
heavy particles.  In addition, if measurements of more than one
super-oblique parameter may be made, some understanding of the
relative splittings in the heavy sector may be gained.
Inconsistencies among the measured values of the different
super-oblique parameters could also point to additional inaccessible
exotic particles with highly split multiplets that are not in complete
representations of a grand-unified group. In addition, negative values
of the parameters will imply new strong Yukawa interactions involving
the SM fields~\cite{CFP}.

If, on the other hand, all superpartners of the standard model
particles are found, the consistency of all super-oblique parameters
with zero will be an important check of the supersymmetric model with
minimal field content.  If instead deviations of the super-oblique
parameters from zero are found, such measurements will provide
exciting evidence for new exotic sectors with highly split multiplets
not far from the weak scale~\cite{CFP}.  These insights could also play
an important role in evaluating future proposals for colliders with
even higher energies, such as the muon collider or higher energy
hadron machines.

In summary, if supersymmetry is discovered, the super-oblique
parameters may allow powerful constraints from precision measurements
on otherwise inaccessible physics.  Their measurement may also have
wide implications for theories beyond the minimal supersymmetric
standard model, just as the oblique corrections of the standard model
provide strong constraints on technicolor models and other extensions
of the standard model.

\acknowledgements

The authors are grateful to M.~Peskin and X.~Tata for conversations.
This work was supported in part by the Director, Office of Energy
Research, Office of High Energy and Nuclear Physics, Division of High
Energy Physics of the U.S.  Department of Energy under Contracts
DE--AC03--76SF00098 and DE--AC02--76CH03000, and in part by the NSF
under grants PHY--95--14797 and PHY--94--23002.  J.L.F. is supported
by a Miller Institute Research Fellowship and thanks the high energy
theory group at Rutgers University for its hospitality.

While completing this work, we learned of related work in
progress~\cite{otherwork}.  We thank D.~Pierce, L.~Randall, and
S.~Thomas for conversations and for bringing this work to our
attention.


\noindent

\noindent

\noindent

\noindent

\noindent

\noindent

\noindent

\noindent

\noindent

\noindent

\noindent

\noindent

\noindent

\noindent

\noindent

\noindent

\noindent

\end{document}